\documentclass[sigconf,balance=false]{acmart}
\usepackage{popets}
\setcopyright{popets}
\copyrightyear{2026}
\acmYear{2026}
\acmVolume{2026}
\acmNumber{X}
\acmDOI{XXXXXXX.XXXXXXX}
\acmISBN{}
\acmConference{Proceedings on Privacy Enhancing Technologies}
\usepackage{tabularx}
\usepackage{multirow}
\usepackage{diagbox}
\usepackage{enumitem}
\usepackage{amsmath}
\usepackage{bbm}
\usepackage{pifont}
\usepackage{hyperref}
\usepackage{xurl}
\usepackage{graphicx}
\usepackage{subcaption}
\usepackage{url}
\usepackage{hyperref}
\usepackage{tikz}
\usepackage{xcolor}
\usepackage{enumitem}

\newcommand{\bcircled}[3][black]{%
  \tikz[baseline=(char.base)]{%
    \node[circle,
          fill=#1,
          text=#2,
          draw=none,
          inner sep=1pt,
          minimum size=1em,
          text width=1em,
          align=center,
          font=\sffamily\bfseries,
          text height=1.5ex,
          text depth=.25ex,
          yshift=0.15ex] (char) {#3};%
  }%
}

\newcommand{\PaperName} {{\it Di5Guise}}
\newcommand{\tamara}[1]{{\color{black}{#1}}}
\newcommand{\camera}[1]{{\color{black}{#1}}}
\begin{document}

\date{}

%





\title{\Large \bf Di5Guise: 5G Privacy with vSIM}

\author{Shirin Ebadi}
\orcid{}
\affiliation{%
  \institution{University of Colorado Boulder}
  \city{} 
  \state{} 
  \country{} 
}
\email{shirin.ebadi@colorado.edu}

\author{Zach Moolman}
\orcid{}
\affiliation{%
  \institution{University of Colorado Boulder}
  \city{} 
  \state{} 
  \country{} 
}
\email{zach.moolman@colorado.edu}

\author{Tamara Lehman}
\orcid{}
\affiliation{%
  \institution{University of Colorado Boulder}
  \city{} 
  \state{} 
  \country{} 
}
\email{tamara.lehman@colorado.edu}

\author{Eric Keller}
\orcid{}
\affiliation{%
  \institution{University of Colorado Boulder}
  \city{} 
  \state{} 
  \country{} 
}
\email{eric.keller@colorado.edu}
\begin{abstract}
SIM cards have been the key building block of user authentication and security in cellular networks. While they are meant to serve as privacy protecting elements in cellular communications, they can be the root cause of privacy loss. Current eSIMs come with a fixed device profile---comprising a secret key, a certificate, and a unique eUICC identifier---that permanently binds every subscriber profile provisioned on the device to that device profile. This binding enables an attacker with the vantage point of a cellular operator to correlate subscriber identities back to a single device, piecing together a complete pattern of life---online activities, movement patterns, and real-world identity---even when users rotate subscriber identities or employ traffic obfuscation techniques.

To mitigate this concern, we introduce \PaperName, 
a privacy-enhancing architecture that breaks this correlation at its root by decoupling the device identity from the subscriber identity. Central to \PaperName~is vSIM, a virtualized SIM card that enables dynamic device profile provisioning, allowing each subscriber profile to be associated with a distinct, unlinkable device profile. \PaperName~establishes trust with the operator by ensuring that vSIM is running on secure hardware in a trustworthy state. We prototype \PaperName~ on a Field Programmable Gate Array (FPGA) board and integrate it with srsRAN to demonstrate full compatibility with existing 5G infrastructure. Using a complex user correlation model, we show that \PaperName~reduces user re-identification accuracy from 93\% to 49\% when combined with obfuscation.

\end{abstract}


\maketitle

%

\section{Introduction}
\label{sec:intro}

Network privacy has long been a desirable goal in research~\cite{wolff2023department,penders2004privacy, alshalan2016vpnsurvey, dingledine2004tor,enck2014taintdroid}. With growing awareness among the general population, there is a growing demand for privacy enhancing technologies~\cite{vsim, bergengruen2022battle,buchanan2020hacker,economistsalttyphoon,mongkolluksamee2016combining,tschimben2024military}. These privacy needs are especially important given that smart phones pervade every aspect of our lives, from communication, to entertainment, to banking. In particular, cellular communication (as opposed to wired) presents a unique privacy concern in that in addition to online activity, physical activity (e.g., in the form of location tracking) is also vulnerable~\cite{lee2023opportunities,huffman2021identifying,minch2004privacy}. The entities best positioned to exploit this information are Mobile Network Operators (MNOs) themselves, who occupy a privileged vantage point with access to subscriber identities, device identifiers, traffic metadata, and location data.

This threat is not hypothetical. Investigative reporting and regulatory actions have documented that major U.S. mobile network operators---including AT\&T, Verizon, T-Mobile, and Sprint---sold customers' real-time location data to hundreds of third-party entities without meaningful consent~\cite{cox2019, whittaker2018}. These carriers offloaded consent obligations onto downstream data aggregators, who in turn resold access to bounty hunters, bail bondsmen, and unauthorized law enforcement users. In 2024, the U.S. Federal Communications Commission levied nearly \$200 million in combined fines against the four carriers for violating federal privacy rules, finding that AT\&T alone shared location data with at least 88 third parties~\cite{krebs}. A federal appeals court subsequently upheld T-Mobile's \$92 million share of those fines~\cite{troy}. 


\if{0}
\begin{figure}[t]
    \centering
    \subfloat[]{%
        \includegraphics[width=.45\columnwidth]{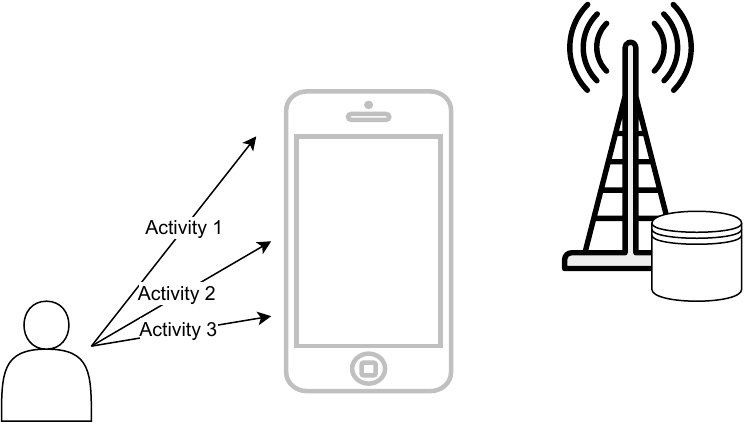}%

        }%
        \hspace{0.1in}
    \subfloat[]{%
        \includegraphics[width=0.45\columnwidth]{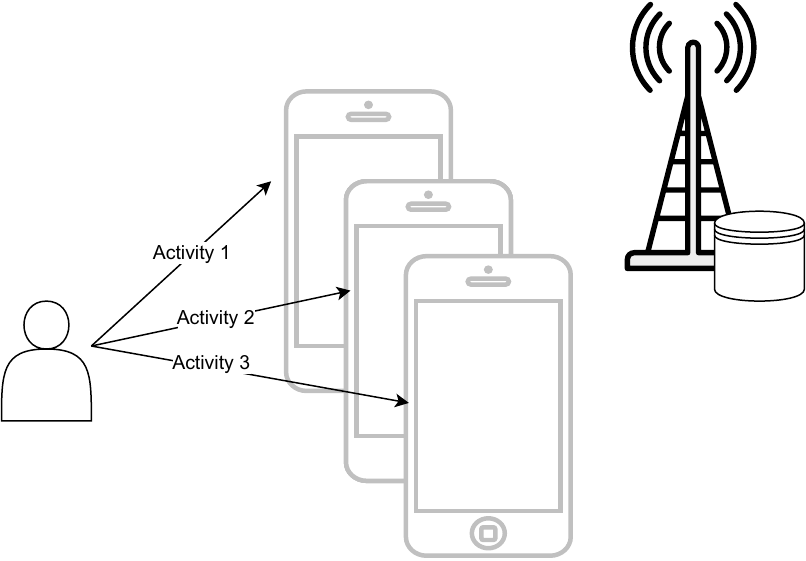}%
        }%
  
    \caption{}
            \label{fig:3vs1}
                
\end{figure}
\fi

We argue that a root of the problem is around the key underlying mechanism for subscriber identification and authentication---the Subscriber Identity Module (SIM) card~\cite{vedder2001subscriber,debaille1997equal}.  A SIM card is a physical chip that stores two pieces of information.  
The first is the subscriber information, which identifies the services the user of the phone should receive.  
In early versions, this subscriber information was read only, prompting the need to swap out SIM cards to change carriers or use a different cellular plan.   
More recently, with eSIM~\cite{gerpott2017embedded,gsm2011embedded}, the SIM card has evolved to enable the ability to dynamically provision the subscriber profile.
The second piece of information stored on a SIM (or eSIM) card is a device identifier, the eUICC Identifier (EID), a certificate, and a cryptographic key, which we collectively call the device profile. This information is fixed at manufacture time and is used to authenticate a device in order to provision new subscriber profiles onto it or manage the subscriber profiles already provisioned on that device.
However, this architecture reveals a significant privacy vulnerability: all subscriber profiles provisioned to a device are tied back to a single, permanent device identifier.

This permanent device binding enables an adversary with the vantage point of a network operator to correlate and link subscriber identities back to a single device.
Even when a user maintains separate subscriber profiles for different purposes, the shared device profile allows the operator to link them, piecing together fragments of an individual's online activities, movement patterns, and real-world identity into a complete pattern of life. Recent works have attempted to hide location and activity from the operator by frequently rotating or anonymizing subscriber identities. 
However, these defenses leave the device identity untouched---the missing piece without which these protections remain ineffective, as the operator can still correlate all profiles back to a single device.
The device profile serves as the root linking point: even a careful individual who maintains multiple subscriber profiles and employs obfuscation techniques (such as a VPN) remains vulnerable. 

To solve this issue, we need to decouple the device from the identity. Today, this decoupling is only possible through physical means such as using different physical phones, manually replacing SIM cards, or attaching a SIM Box which houses a pool of SIM cards~\cite{glocalme}. However, these approaches are cumbersome, impractical, and do not scale. To efficiently and scalably address this problem, we propose a novel SIM architecture that moves away from fixed physical secure elements and instead introduces a virtualized SIM card, called \textit{vSIM}. vSIM allows users to remotely provision device profiles and seamlessly switch between them, so that each subscriber profile can be tied to a completely different device profile---enabling a user to maintain multiple unlinkable identities on 5G networks.

In moving to an infrastructure that utilizes a virtualized SIM card, a number of challenges are introduced when moving away from implicit assumptions of the properties of today's physical SIM card. Without a dedicated hardware boundary, the secrecy of sensitive credentials is no longer inherently enforced; the operator's trust model, built around immutable manufacturing-time credentials, no longer applies when these can be provisioned dynamically; and the operator needs assurance that an anonymous device requesting a new profile is running a secure software on a trustworthy platform that can protect secrets. Moreover, the current infrastructure assumes a permanent device credential and fixed EID for subscriber profile binding---assumptions that break entirely under dynamic device profile provisioning. To address these challenges, we introduce \PaperName{}, an architecture built around vSIM that proposes novel solutions to establish the trust guarantees that physical SIMs provided through hardware, while enabling the dynamic device profile provisioning needed to break linkability.



\if{0}

\begin{figure}[t]
    \centering
    \includegraphics[width=\linewidth]{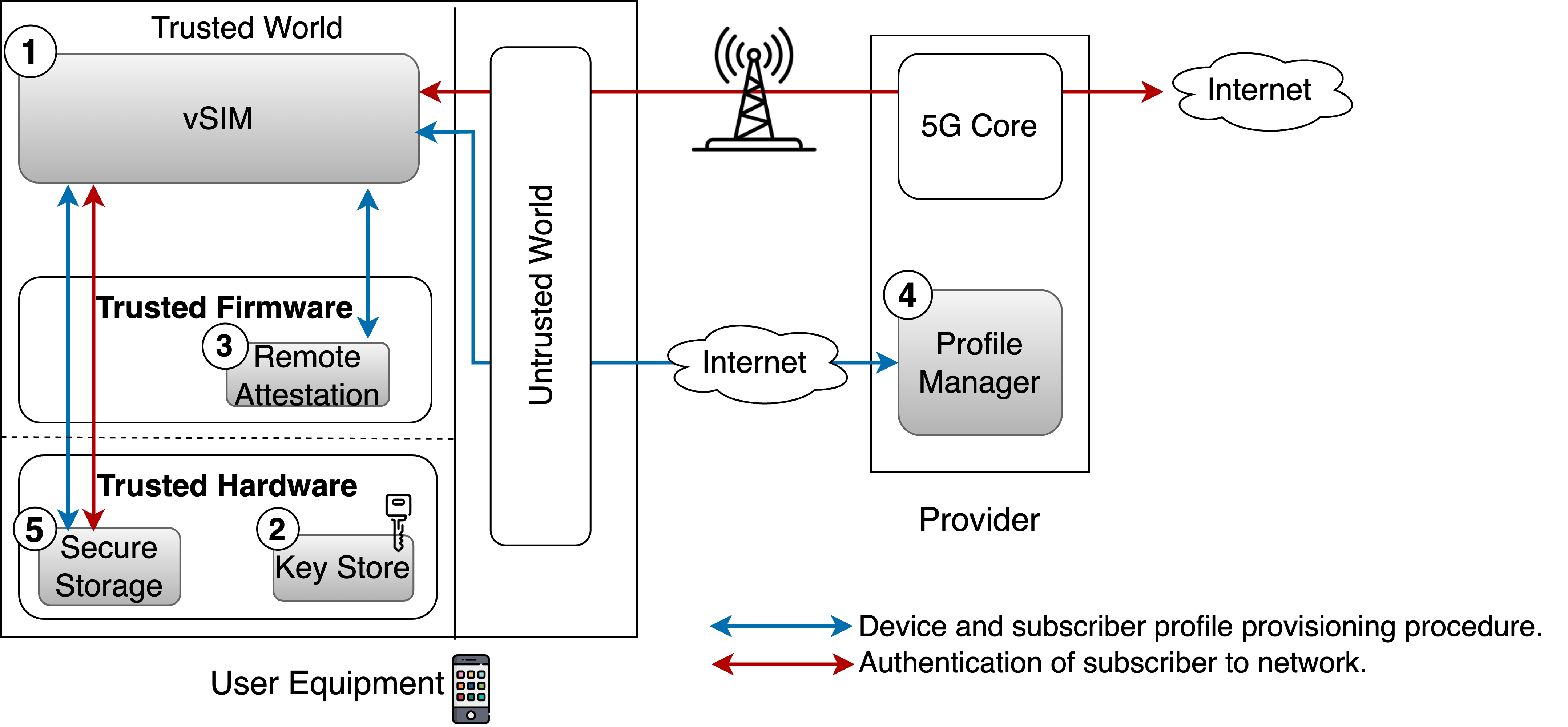}
    \caption{\PaperName{} Overview. 
    (1)  vSIM which provides hardware-enforced isolated SIM card functions in software to provision profiles dynamically and privately,
(2) the trusted hardware,
(3) the remote attestation,
(4) the Profile Manager, and
(5) the successful establishment of a profile in secure storage.}
    \label{fig:overview}
\end{figure}

\fi

We prototype \PaperName{} on a Xilinx Genesys2 FPGA board. We modify the Software Radio System Radio Access Network (srsRAN) User Equipment (UE) software (mobile equipment side) to use our design of vSIM for 5G authentication instead of eSIM.  Our evaluation shows that
\PaperName{} greatly increases privacy.  We evaluate the effectiveness using SiamHAN~\cite{cui2021siamhan}, a machine learning model used to correlate network traffic patterns originating from seemingly different users.  Compared to a baseline of eSIM plus TLS obfuscation, with \PaperName{}, SiamHAN's accuracy drops from 93\% down to 49\% accuracy, which, given the distribution of the dataset, is no more effective than randomly guessing.


Our contributions can be summarized as follows:
\begin{itemize}[nosep]
  \item Introducing a new SIM architecture that is able to provide functionality needed for 5G authentication, with the added benefit of dynamic device profile provisioning.
  \item Providing an architecture which leverages several techniques such as trusted execution, secure hardware, remote attestation, and group keys, to provide a means to establish trust in the absence of a trusted physical SIM.  
  \item A fully working prototype that integrates with open source 5G software, along with an evaluation that demonstrates significant privacy gains with small overhead.
\end{itemize}

\tamara{In this section, we motivate our work by discussing background on 5G infrastructure and how it is not sufficient in maintaining an individual's privacy. An individual's privacy is impacted by an attacker that can correlate multiple subscriber profiles to track an individual's activity---online (when and what) and physical location (where) activity~\cite{shklovski2018use,horbyk2022war,saad2018security}\footnote{Cases where the adversary is physically following a user (tailing) is out of scope.}. We then overview the specific threat model we are targeting.}

\if{0}
\subsection{Traffic Obfuscation / De-obfuscation}

Even with encrypted traffic, a network operator can still gain a lot of information about a user. We generalize to any entity with which is used to access the Internet, whether wired or wireless. As a general rule, IP addresses are known.  The source IP address is usually determined by the network operator dynamically, but can be tracked across authenticated sessions, if desired.  The destination address, however, can be used to identify activity, such as which websites are visited.  This information can then be used to form a profile of a user~\cite{gonzalez2016userprofiling,tschimben2023modeling}.

A popular approach to work around tracking sites visited is to use a tunnel.  One example of this approach is a virtual private network (VPN)~\cite{alshalan2016vpnsurvey}, where a VPN operator creates a tunneled connection between the client and its server (hiding the destination IP address in an inner packet, where only the destination IP address of the VPN server is shown).  Another approach is onion routing (such as with Tor~\cite{dingledine2004tor}), which tunnels traffic to an ingress network node (among many choices), and then bounces around various intermediate nodes before being forwarded to the destination (with an array of wrapper headers to hide the source and destination).

However, many approaches exists to bypass obfuscation techniques and identify users using other pieces of information. For example, in the case of Tor, an attacker can correlate ingress and egress traffic to de-anonymize the source and destination~\cite{KarunanayakeIEEEcsur2021, deepcorr}.  Other techniques for de-obfuscating users include fingerprinting activity patterns~\cite{tschimben2024military, van2020flowprint}, characterizing unique properties of the Operating System (OS)~\cite{os-finger}, or the device~\cite{oh-ndss2023}, or the protocol~\cite{sni-finger}. User behavior can be used to identify specific instances, which can then be used to correlate different flows and associate traffic to specific users~\cite{cui2021siamhan, useripprofiling}.

In response, users can obfuscate their traffic to hide these properties and increase their privacy.   Techniques vary from randomizing traffic~\cite{obfs4, scramblesuit, tor-pt}, mimicing characteristics of a known protocol~\cite{stegotorus, skypemorph}, and padding traffic~\cite{meier2022ditto, buflo, csbuflo, hornet, taranet}.  In each case, these techniques make traffic analysis much more challenging.   
\fi

\if{0}
\begin{figure}[t]
    \centering
    \includegraphics[width=0.9\linewidth]{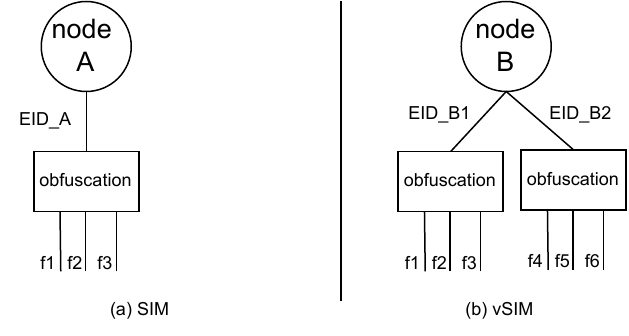}
    \caption{(a) shows a node A using eSIM with a fixed EID, EID\_A.  Obfuscated traffic flows f1, f2, and f3 will all be associated with EID\_A.  (b) shows nodes using vSIM with dynamically provision EIDs (and associated eSIM secret keys), EID\_B1 and EID\_B2.  In this case, flows (f1, f2, f3) and flows (f4, f5, f6) will look like they are coming from different nodes.}
    \label{fig:nodes}    
\end{figure}
\fi
\section{Background and Motivation}

\subsection{5G Network Architecture Background}

\subsubsection{SIM Card Architecture and Application}

The Universal Integrated Circuit Card (UICC), also casually referred to as a SIM card, is a smart card used in cellular infrastructure~\cite{ETSI_TR_102_216}.
They can come as a removable or non-removable embedded hardware module, 
with a microprocessor from manufacturers such as Intel, Motorola, or STMicroelectronics, and up to 256 KB of volatile memory~\cite{8554774}. \tamara{The UICC has evolved from a physically insertable device (\emph{i.e.} SIM card), to a device soldered on the system board (\emph{i.e.} eSIM), to one that is embedded in the system on a chip (\emph{i.e.} iSIM).}

\tamara{Within the UICC runs the Universal Subscriber Identity Module (USIM), which is an application that services the transactions required in mobile communications} in Universal Mobile Telecommunications Service (UMTS or 3G), Long-Term Evolution (LTE or 4G), and 5G devices. 
\tamara{USIM hosts several fundamental services required for mobile communications and performs security functions like ciphering and authentication of the user.}  



\begin{figure}
    \centering
    \includegraphics[width=\columnwidth]{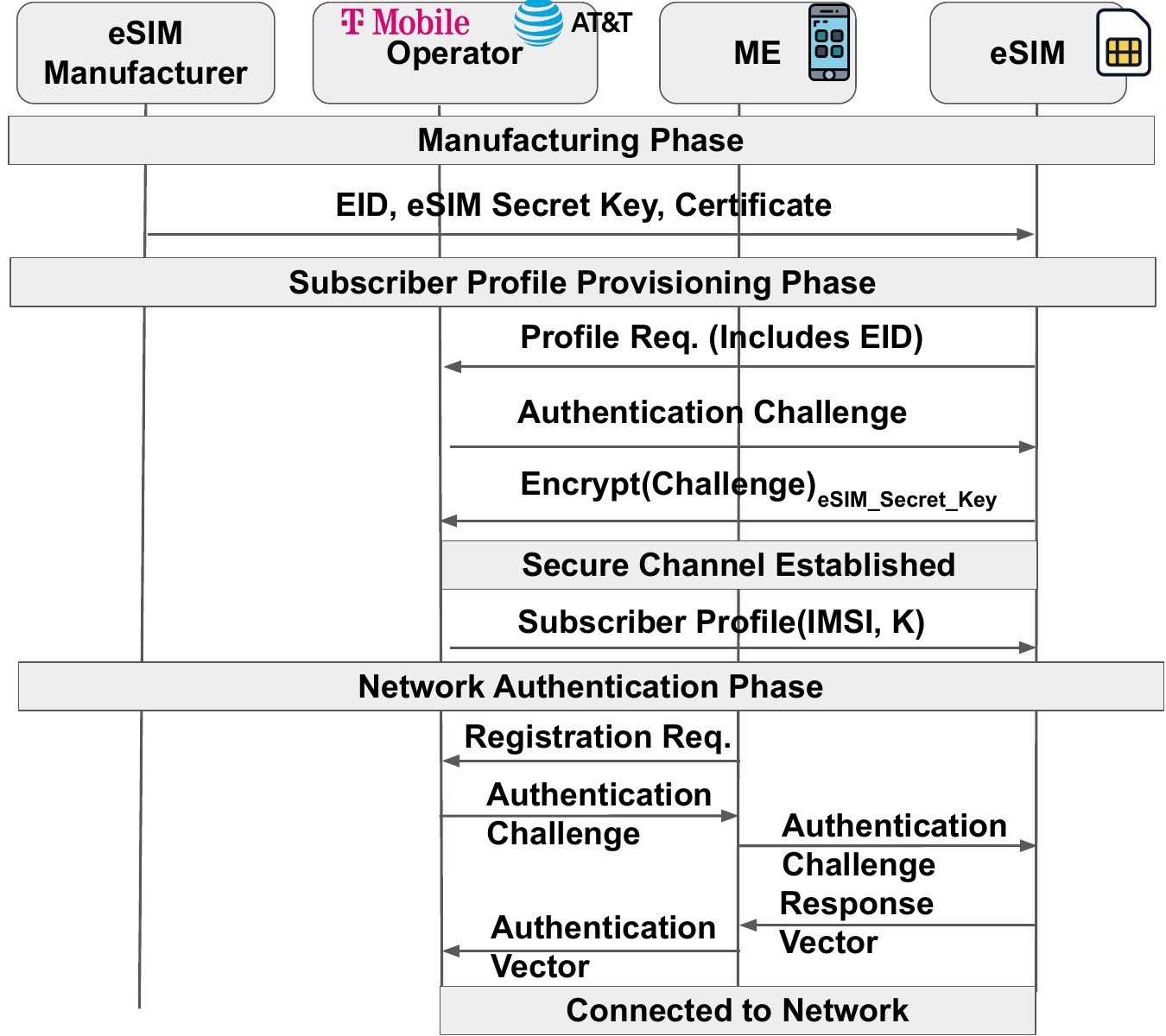}
    \vspace{-2em}
    \caption{Steps for a SIM Card to connect to the network. }
    \label{fig:auth}
\vspace{-1.5em}
\end{figure}

Historically, a UICC was pre-provisioned once during manufacturing with a subscriber profile. \tamara{The UICC includes an International Mobile Subscriber Identity (IMSI), which is a unique 15-digit number used by MNOs to identify mobile subscribers when they connect to a cellular network. The IMSI is created and shared by the mobile operator.} To address the inflexibility of this method, the Global System for Mobile Association (GSMA) specified a new protocol that can remotely provision a new subscriber profile into an embedded UICC (eUICC). Figure \ref{fig:auth} illustrates the simplified phases through which a genuine eUICC provisions a new profile and authenticates to the network. For this use, at manufacturing time, a secret key (eSIM secret key) is provisioned into the eUICC. Additionally, each device comes with a globally unique Device Identifier, the eUICC identifier (EID).  The secret key is not readable, but the device identifier is. Later, these credentials are used to authenticate eUICC to the operator in order to provision a subscriber profile used to authenticate to the network.

For example, to activate Google Fi service on a mobile phone with an eSIM, the customer needs to download the Fi application. 
The Google Fi app will interface with Google servers (over Wi-Fi at this point) and request a new subscriber profile. After authenticating the eUICC using the device identifier and pre-shared eSIM secret key, a new subscriber profile will be encrypted and sent through a secure channel to the eUICC. The eUICC will then install the profile, enabling the device to connect to the Google Fi network.

\label{sec:background}

\subsubsection{Subscriber Profile Provisioning and Authentication Process} \label{sssec:5gauth}
As shown in Figure~\ref{fig:auth}, the eSIM comes with a device profile that is used to authenticate the eUICC. The eSIM needs to provision a subscriber profile in order to connect to the network. It begins the process by sending a request to the operator (SM-DP+) server to prepare a new profile. The eSIM secret key is used to create a digital signature to authenticate the eSIM to the operator. The operator prepares the profile for download.  From the operator side, the generated profile will be linked to the provided EID with for cross-network profile identification and operational management.

After securely downloading the subscriber profile, the user can connect to the network. To do this, the operator needs to authenticate the subscriber profile to its network. In 5G, this is achieved through primary authentication, which is a mutual key exchange process. The 5G standard proposes the Extensible Authentication Protocol Authentication Key Agreement (EAP-AKA) or the 5G-AKA as the authentication protocol~\cite{3gpp_ts_33_501}. The process is initiated by the Mobile Equipment (ME)\footnote{Mobile Equipment (ME) refers to the device and software as a distinct entity from the SIM card. The combination of ME and SIM is the User Equipment (UE).} sending a registration request containing the subscriber identity. Upon reception of the request, the operator validates it and computes a method-specific authentication vector including a random number (RAND) and Authentication Token (AUTN) and sends it to the ME. The ME, in turn, forwards it to the SIM. After validating the received vector, the SIM calculates the response vector and sends it back to the ME, which then sends an authentication response to the operator. The eSIM calculates its responses based on the subscriber long-term secret key (K) using the AES encryption algorithm~\cite{primitives2003advanced}. The SIM software implements several cryptographic functions defined in the standard that are used, directly or combined, to calculate the response vector~\cite{3GPP:TS31.102:R18}.

Once authenticated, the eSIM application is no longer involved in handling data traffic. The IMSI (the unique 15 digit number) serves as the subscriber identifier, which mobile network operators use to determine service provisioning.
To address security vulnerabilities, 5G introduced an additional layer of protection against IMSI-catcher attacks~\cite{imsi-catcher}. In 5G networks, the IMSI---generically referred to as the Subscriber Permanent Identifier (SUPI)---is concealed through encryption using the operator's public key, creating the Subscription Concealed Identifier (SUCI)~\cite{3GPP:TS31.102:R18}.


\subsection{Pattern-of-Life Analysis in 5G}
\label{motivation:pattern}

\tamara{The current 5G infrastructure binds together the device and the user's profile creating privacy concerns.} The SIM/eSIM is provisioned at manufacturing time with a unique device identifier (EID), a certificate, and a secret key (which, together we call the device profile). The operator ties this device profile with each subscriber profile provisioned to that device. This binding creates a tight connection between the user's network activity and a specific device. 

\noindent\textbf{Honest-but-Curious Operator.} Recent studies have reported that many operators collect and sell user information to third-party companies~\cite{krebs,troy}. Operators occupy a privileged vantage point, as they have access to a wide range of user information, including subscriber identities, device identifiers, location data, and traffic metadata. This enables an adversary with some level of control over the operator---whether through a compromised insider, a commercial data-sharing arrangement, or state-level compulsion---to readily build detailed profiles of individual users. 

Permanent identifiers, whether at the hardware level or the subscriber level, are the main linking points~\cite{mbtrack}. With a permanent device profile (which consists of an EID and a security key), an operator can easily link subscribers---even those purchased under different real-world identities for different purposes---to the same device profile~\cite{esim}. This means that all online activities of an individual, along with the ``what, when, and where'' are associated with the same device regardless of which subscriber profile is in use.

This link is problematic for several reasons. First, 
it enables the operator to cluster subscribers based on their device profiles and use the activity across different subscriber profiles on the same device---each of which may serve a different purpose---to complete the puzzle of a user's pattern of life. Second,
a subscriber profile is easily trackable even when disabled, based on the activity of another enabled profile on the same eSIM, even if that profile was registered under a different real-world identity.



\begin{table}[t]
\centering
\caption{Comparison of defenses against operator-side attacks on subscriber and device privacy. A \ding{51} indicates that the referenced tracking/correlation is prevented and a \ding{55} indicates that it is still possible.}
\vspace{-1em}
\label{tab:works}
\footnotesize
\setlength{\tabcolsep}{2pt}
\begin{tabular}{c|c|c|c|c}
\hline
 & \textbf{\textit{Subscriber Tracking}}  & \multicolumn{2}{c|}{\textbf{\textit{Device Tracking}}}  & \textbf{\textit{Cross-Subscriber}}\\
 & \textbf{\textit{(SUPI, Location)}} & \multicolumn{2}{c|}{}& \textbf{\textit{Correlation}}\\ [-1em]\cline{3-4}
 &   & \textbf{\textit{IMEI}} & \textbf{\textit{(EID, eSIM Key)}}& \\
\hline \hline
\textit{PGPP~\cite{schmitt-usec2021}}  & \ding{51}  & \ding{55} & \ding{55}  & \ding{55}\\
\hline
\textit{AAKA~\cite{yu-ndss2024-aaka}}      & \ding{51}  & \ding{55} & \ding{55}   & \ding{55}\\
\hline    
\textit{ZipPhone~\cite{zipphone}}      & \ding{51}  & \ding{55} & \ding{55}  & \ding{55}\\
\hline
\textit{PPIC~\cite{randomimei}}  & \ding{55}  & \ding{51} & \ding{55}  & \ding{55}\\
\hline
\textit{ADA~\cite{pei}}    & \ding{55}  & \ding{51} & \ding{55}  & \ding{55}\\
\hline
\textit{\PaperName{}}       & \ding{55}  & \ding{55} & \ding{51}& \ding{51} \\
\hline
\end{tabular}
\vspace{-2em}
\end{table}

Most prior works focus on defending against subscriber-level attacks from a passive or active outsider attacker. More recently, several solutions have been proposed to defend against the operator itself. These works, summarized in Table~\ref{tab:works}, address subscriber-level tracking by anonymizing the subscriber identity (IMSI, or more generally the SUPI in 5G). They tackle subscriber privacy from different approaches: decoupling authentication from connectivity~\cite{schmitt-usec2021}, making authentication sessions unlinkable through anonymous credentials~\cite{yu-ndss2024-aaka}, and rotating subscriber identities among a community of users to frustrate location profiling~\cite{zipphone}. However, these solutions remain incomplete in the presence of a permanent device identifier, as the underlying infrastructure couples all subscriber profiles to the same device the moment it is generated---even before any subscriber profile is activated. In fact, any subscriber-level solution that uses ephemeral subscriber identifiers and rotates among them is completely defeatable through EID tracking across identifier changes. Even in works that fully rely on provisioning anonymous credentials or dummy SUPIs, a critical piece remains unaddressed: how profile management operations (\S\ref{sssec:5gauth}) are handled.

Profile management fundamentally requires the eSIM to authenticate using its device profile to operator and to manage profiles indexed by their subscriber identifiers. This inevitably preserves a binding between the EID and the associated subscriber identifiers. Since multiple IMSIs are provisioned under a single unique EID, the infrastructure enforces a structural one-to-many mapping from a unique EID to multiple unique IMSIs. 

Complementary to subscriber-level defenses, recent work has addressed the privacy of the IMEI---referred to as the Permanent Equipment Identifier (PEI) in 5G---which the core network can request after successful authentication~\cite{randomimei, pei}. While these approaches conceal the IMEI during equipment identity verification, they still leave the underlying identifier untouched. Therefore, fully achieving untrackable and unlinkable connectivity requires first systematically decoupling the device from subscriber identity.

\tamara{\PaperName{}'s main goal is on breaking the linkability across subscriber profiles by decoupling identity from the underlying device context. \PaperName{} does not aim to solve subscriber-level identity or location tracking in isolation, both of which have been done before. Subscriber-level and IMEI anonymity mechanisms (Table~\ref{tab:works}) are complementary to \PaperName{} and can be integrated to provide stronger, layered protection against a curious operator. }
\subsection{Threat Model}

The threat model considers the use case where an individual has a smart phone connecting through a commercial 5G network with a paid relationship with the cellular operator.  This relationship can be in the form of pre-paid accounts which do not require personal IDs for a consumer to acquire~\cite{burner, pre-paid}. \tamara{We acknowledge the fact that in certain regions this option may not be possible without identifying the user. 
}We assume the attacker has the vantage point of the cellular operator, where the operator is honest but curious (tracking its users for commercial use~\cite{cox2019, whittaker2018}) or compelled by a state actor to provide access for the network access points in that country~\cite{economistsalttyphoon,287260}.  
In the latter case, the attacker can have the vantage point of multiple cellular operators (again, restricted to the access points in the jurisdiction of the state actor).
We assume the metadata is fully accessible by the adversary. This includes any unique identifiers needed for eSIM authentication and subscriber authentication, and features of network traffic.
Our focus, however, is on the device identifier, as prior works already address subscriber identity and location privacy.

\section{\PaperName{} Overview}
\label{sec:overview}


Solving the correlation problem requires the ability to dynamically provision device profiles---a capability that current physical SIMs, by design, cannot provide due to their hardware-bound credentials. By moving SIM functionality into software (virtualized SIM), device profiles are no longer fixed---they become programmable and can be provisioned, replaced, and switched on demand. 

However, breaking correlation through virtualizing the SIM and dynamically provisioning a device profile introduces a number of challenges due to implicit assumptions the 5G infrastructure makes about the physical properties of a SIM card. These challenges necessitate the design of \PaperName{}, an architecture built around vSIM. Here, we overview challenges and how \PaperName{} solves them.

\textbf{Challenge 1: SIM key secrecy.} In the current infrastructure, a SIM (in all current forms) is a fixed secure element, which has hardware-enforced protection of the manufacturer-burned secret credentials (ID and secret keys). So basically the current secure element hardware is the main root of linkabality and not fundamentally capable of supporting any changes in it.  

To address this challenge we design the virtualized SIM (vSIM), a software version of the SIM functionality that runs in a hardware-enforced isolation mechanism available in many modern processors~\cite{intelsgx,alves2004trustzone,lee2019keystone}---a trusted execution environment (TEE) within the CPU of a mobile device.  \tamara{The TEE provides an isolated execution environment in which not even the operating system is able to extract secret data. In addition, several TEE implementations support secure memory, which provides an additional layer of privacy and security, preventing an external observer with physical control of the device 
from accessing secrets.}

\textbf{Challenge 2: Establishing initial trust with the SIM card.} In the current 5G infrastructure, trust is rooted in a business relationship between the manufacturers and operators. As part of this trust, the operators trust any device that can authenticate using the credentials provisioned at manufacturing time. The operators allow this trust because the manufacturers, through established business processes and certification requirements, ensure that the device and its secure element are implemented correctly and securely. With a virtualized (software) SIM (the foundation for \PaperName{}) where the secret key and certificate can be provisioned dynamically, the operator can no longer rely just on that business relationship.  

\tamara{We address this challenge by leveraging the concept of hardware-based group cryptographic keys. The group keys are used to establish trust in the hardware while preserving the individual device anonymity. We combine the group keys with a secure boot process to establish the initial trust of the secure firmware running on the device. These two components together (group keys and secure boot) provide the basis for establishing trust without relying on an individual hardware key. Once this initial trust is established, a new secret key can be provisioned dynamically to become what used to be the secret key stored on the physical SIM card, except that now it can be dynamically provisioned without revealing the device identity.}

\textbf{Challenge 3: Trusting the dynamic device profile provisioning.} In the current 5G infrastructure, the device profile (device identity, certificate, and secret key) is fixed at manufacturing time and communicated to the operator through established business processes. With vSIM, the device is dynamically provisioned as needed by the privacy model. This dynamic provisioning is the key novel capability of the proposed design but it introduces yet another challenge on trust. \tamara{Given that the dynamic provisioning needs to be established by an anonymous entity (i.e. the mobile device), the profile provisioner (i.e. the operator) needs to have some certainty that the software running on the device is secure and trusted.}

To address this challenge, \PaperName{} introduces a protocol to achieve this provisioning securely, which starts with performing remote attestation to verify and authenticate the hardware and software running in the TEE. The vSIM software establishes a secure channel (running in a secure enclave in the TEE), and finally provisions the device profile.

\textbf{Challenge 4: Subscriber profile provisioning with a dynamically provisioned device profile.}
The current 5G infrastructure assumes a permanent device credential—a secret key and certificate provisioned at manufacturing time—used to authenticate the eUICC to the operator, along with a fixed EID used to generate and bind subscriber profiles to that device. With vSIM, all are provisioned dynamically, so this permanent binding no longer holds.

We construct the vSIM software to have an interface that is entirely backwards compatible. With this flexibility, the practicality of deployment is quite high---the current SIM manufacturers or even a new third party (which establishes a business relationship with the operators) can serve the role of device profile provisioning, and once that is done, all of the 5G infrastructure works with no modifications. That is, vSIM still uses a secret key to authenticate and provides an EID for subscriber profile generation and binding---the difference is that instead of a single permanent identifier and secret key, the user can have multiple, each used with a different subscriber profile. As a result, all steps from generating a new subscriber profile, to provisioning it on the device, to authenticating to the network remain unchanged.

\section{\PaperName{} Architecture}
\begin{figure*} 
    \centering
    \includegraphics[width=0.95\textwidth]{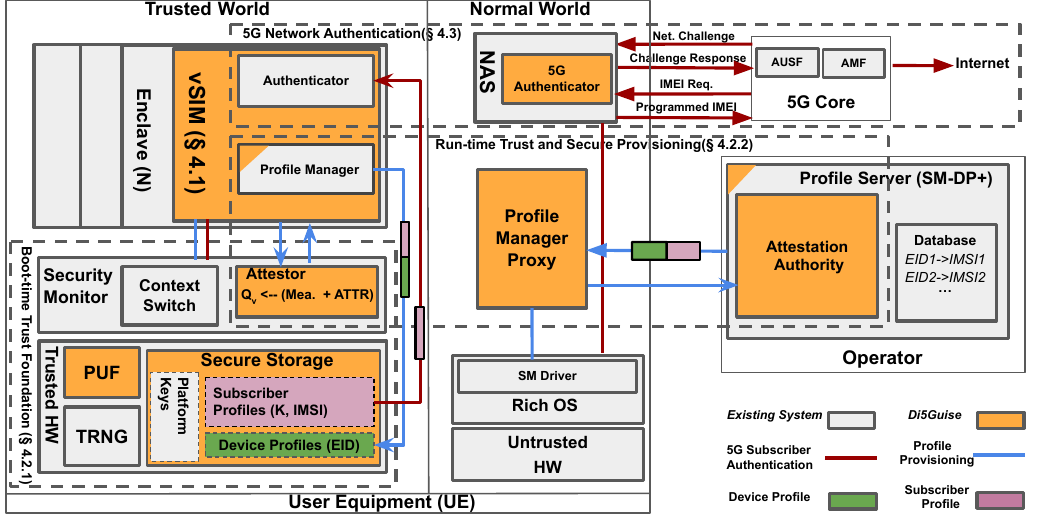 }
    \vspace{-1em}
    \caption{Architecture of \PaperName{} connected to 5G core network. }
    \label{fig:arch}
\end{figure*}

In this Section, we detail the architecture of \PaperName{}, illustrated in Figure~\ref{fig:arch}.
At its core, \PaperName{} includes two key innovations. 
First is the vSIM software (Section~\ref{sec:arch:vsim}).
Second, we introduce supporting capabilities that enable \PaperName{} to be trusted and the dynamic device profile provisioning  (Section~\ref{sec:arch:trustest}), which is the unique capability that allows a user to use different device profiles and therefore reduces an adversary's ability to track the user's activity.
After detailing each of these, we discuss the runtime operations (Section~\ref{sec:arch:runtime}), to illustrate how 5G can work without modifications.

\subsection{vSIM: Virtualizing the SIM}
\label{sec:arch:vsim}

vSIM is a software-based implementation of SIM functionality.  
Like SIM and eSIM, it includes both a device and subscriber profile and  supports 5G authentication.  Where vSIM differs is \textit{\textbf{(i)}} it is software running on the main processor, not a physically separated device, and \textit{\textbf{(ii)}} it is able to dynamically provision the device profile (including EID).   

In order to isolate the execution of the vSIM from other system software,
vSIM is designed to run in a trusted execution environment
(TEE). In traditional processors, an Operating System (OS) has
complete control over the system. It can, for example, inspect and modify any memory, including memory used by applications. This ability poses a threat as the information used in a
SIM card is intended to not be readable by the system. \tamara{A TEE
is a hardware construct that provides the ability to execute software in an isolated context away from the OS control and enforced by the hardware. TEE's are supported by most modern processors (RISC-V Keystone~\cite{lee2019keystone}, ARM TrustZone~\cite{alves2004trustzone}, Intel SGX~\cite{intelsgx}, Intel TDX~\cite{inteltdx}, etc.).}

\tamara{This isolation is enforced by leveraging a new mode of operation: the machine mode \footnote{The description of the TEE presented here is based on the RISC-V Keystone~\cite{lee2019keystone} version but the main concepts apply to most other TEE architectures.}}. This mode is used to run software (firmware), called the security monitor, that is verified before it is loaded, \tamara{during the secure boot process,} to ensure its authenticity and functionality. Unlike the OS, the security monitor is small enough to allow its functionality to be easily verified and authenticated during the boot process. The security monitor takes advantage of hardware structures only available in this mode of execution to manage memory access permissions and context switching. The OS runs in the traditional privileged mode, but it does not have access to the hardware structures that require machine mode. \tamara{Leveraging the memory access permission hardware structure, the security monitor can isolate the memory for a particular TEE so that it is separated from all other memory allocations in the system, including the OS. }

The vSIM software is executed in a secure enclave of a TEE.  \tamara{The interface used by the software to control the interactions between the trusted and untrusted worlds are done using \textit{ECALLs} and \textit{OCALLs}. \textit{ECALLs} are a narrow and well-defined interface to call trusted functions from the untrusted application code. \textit{OCALLs} are used to call functions from the trusted code back into the untrusted application. Parameters in both of these interfaces are passed as shared memory regions.} The communication occurs through the Security Monitor (in hardware mode) via the Security Monitor driver within the OS. 

Figure~\ref{fig:arch} illustrates two main processes supported by vSIM. Unlike the traditional SIM, vSIM also includes the device profile provisioning capability (illustrated by the blue arrow). This interface uses a proxy (Profile Manager Proxy) running in the non-secure world, which interfaces with the profile server in the operator's infrastructure. This proxy calls vSIM to first establish a secure channel and then downloads the device profile with \emph{downloadDeviceProfile()}. This process is discussed in more detail in Section~\ref{sec:arch:trustest}. It also supports subscriber authentication and connection to the network (illustrated by the red arrow). For authentication, vSIM exposes the \emph{challenge()} call, invoked by the 5G Authenticator module running in the non-secure world. This function takes in the \emph{RAND} and \emph{AUTN} parameters of the 5G authentication process, as outlined in Section~\ref{sssec:5gauth}. Upon invocation, the 5G Authenticator module in \PaperName{} context switches into the protected environment, the secure enclave, to complete its operation.


\subsection{Device Profile Provisioning}
\label{sec:arch:trustest}

Central to the privacy enhancements that \PaperName{} enables is the ability to dynamically provision a device profile.  Recall that a device profile consists of a device ID  (EID), a certificate, and a secret key (eSIM secret key). In order for an operator to be comfortable provisioning (and then subsequently authenticating) a device profile to a remote device, it needs to trust that the device is trustworthy and can protect secret information (particularly the eSIM secret key). \PaperName{} enforces this secrecy through multiple mechanisms that establish a foundation for trusting the hardware, firmware, and vSIM software and then builds upon this foundation to establish a secure channel for transferring the device profile.
\subsubsection{Boot-time Trust Foundation}

If an adversary has access to the external device where the firmware is stored, it can compromise the system by changing the firmware.  Being able to trust the security monitor, in particular, is critical to establishing trust with \PaperName{}.


To mitigate this potential vulnerability, \PaperName{} includes a secure boot sequence.  Secure boot is the process of bringing up a system in such a way that it guarantees that once the boot process has been completed, the system has not been compromised in any way and that only authenticated software has been used during the boot process.
\begin{figure} 
    \centering
    \includegraphics[width=\linewidth]{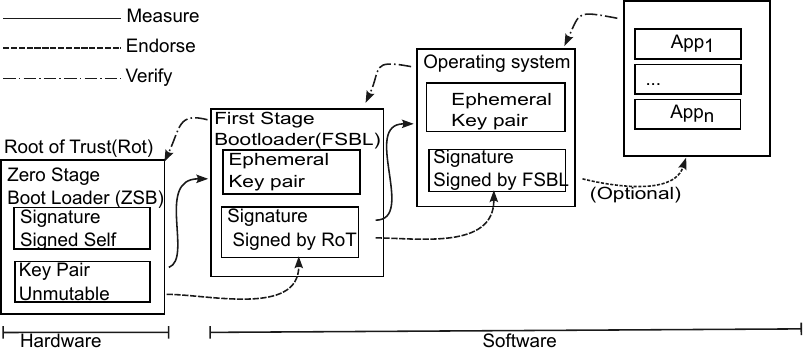}
    \caption{Secure Boot and Chain of Trust.}
    \label{fig:bootsequencecot}
\end{figure}
The boot sequence, illustrated in Figure~\ref{fig:bootsequencecot}, includes a boot vector that is the starting point for each core. The program located at the boot vector is a very small program called the Zero-Stage-Boot-Loader (ZSBL) that is fused into the silicon (BootROM) to start the boot process.  On a hard reset, each core begins executing the instructions from the boot vector.  The main objective of the ZSBL is to load the firmware, which contains the First-Stage Bootloader (FSBL) and some optional data that might be required to complete the boot process.  To make progress, the  ZSBL has to manage all the available cores (each core comes online independently) and initializes the minimum I/O needed to load the firmware stored on an external device (e.g.  SDCard or BIOS) into main memory.  The control is then passed to the First-Stage-Bootloader, which loads the Secondary-Stage-Bootloader or the Operating System.

To ensure the fused-in code (the ZSBL) is the correct code, the device includes a cryptographic key fused into the silicon with which it signs the ZSBL~\cite{maes2012pufky,bhargava2014efficient}. Starting at the ZSBL, each layer measures and endorses the next layer until control is handed over to the operating system.  To measure the next layer, the current trusted layer verifies the next layer with the next layer’s digital signature and public key to ensure that the image has not been modified.  If the measurement fails (the stored signature differs from the computed signature), the boot process halts, preventing the loading and execution of malicious software.  Otherwise, the current layer will endorse the next layer by signing the next layer with its private key before transferring control to it.

\label{sec:arch:profile}
\subsubsection{Run-time Trust and Secure Provisioning}
  While the hardware root of trust and secure boot together establish trust in the hardware and firmware at boot time, the operator also needs assurance at run time that the correct vSIM software is running and that the communication channel is secure. Di5Guise achieves this through two mechanisms: remote attestation, which allows the operator to verify the current state of the platform and the vSIM application, and a secure channel establishment protocol, which ensures that the device profile can be transferred confidentially.



\label{sec:arch:secchan}

\textbf{Remote Attestation. } 
Remote attestation provides a means for a remote party to verify what is currently running on the valid and trusted platform---so it can be sure only the intended software, and no additional applications have been added.  We do not consider the vSIM application to be part of the secure boot process, but something that may have different implementations that could be loaded. The vSIM remote attestation is a system level software that executes in the secure world and is part of the secure boot process.  \PaperName{} needs to prove that \textit{\textbf{(i)}} it is currently running on a valid and trusted platform (secure monitor is trustworthy)---so the operator can be sure that the intended software is running---and  \textit{\textbf{(ii)}} that it is in a valid and unmodified state. In \PaperName{}, multiple components can be remotely attested: \textit{\textbf{(i)}} the physical hardware device, \textit{\textbf{(ii)}} the TEE Secure Monitor, and \textit{\textbf{(iii)}} the vSIM application itself. To this end, \PaperName{} responds to cryptographic challenges from the operator with the attestation quote. The quote includes the hardware state, which includes measurement of the Security Monitor (SM) and its signature, and measurement of vSIM and its signature. All this information is signed by the device's private group key along with the challenge and sent to the remote verifier. The operator verifies the signature to confirm that the response originates from a genuine \PaperName{} hardware (by confirming membership of the group key) and that it and the vSIM software are in valid and trusted states.

  \textbf{Device Private Group Key.}
It is crucial in the design of \PaperName{} that no unique hardware information (e.g., cryptographic keys) is shared with any third party (including the remote authority) to preserve the user's identity.  
However, the operator needs to verify the hardware device integrity to ensure that the software runs on ``genuine \PaperName{} hardware'' and can, therefore, be trusted.

To achieve this goal, we design \PaperName{} to use a group key scheme based on the direct anonymous attestation scheme (DAA)~\cite{10.1145/1030083.1030103}. DAA is based on the idea that a set of devices can be part of a group which can create the same signature without revealing which device in the group created it~\cite{10.5555/1754868.1754897}. Each device has a unique cryptographic public-private key in a traditional digital signature scheme.  
However, in DAA, each device holds a unique private key (the platform private key) that is cryptographically associated with a common group public key (the platform public key) used for verification~\cite{alsouri2010group,chen2022mage,scarlata2018supporting}. DAA has been widely applied in scenarios similar to vSIM, where a device has to prove it is part of the group while preserving its anonymity—for example, Intel’s Enhanced Privacy ID (EPID) scheme~\cite{johnson2016intel}. 

\camera{DAA-based systems such as Intel's EPID also support credential revocation and recovery if they are compromised~\cite{epidiot}. Specifically, \textit{\textbf{(i)}} if an individual device's private key is compromised, the issuer can add the corresponding private key to a revocation list. \textit{\textbf{(ii)}} if a compromised or malicious signature is detected, the signature can be added to a signature-based revocation list. \textit{\textbf{(iii)}} in the more severe case where the group credential is compromised, the issuer can revoke the entire group and provision a new group public key and member credentials.}

  \begin{table}
  
\caption{Protocol notations}
\vspace{-1em}
\label{tab:notation}

    \centering
 \begin{tabular}{ll}
        \hline
        \textbf{Notation} & \textbf{Description} \\
        \hline
         $n_X$ &  Nonce generated by party X \\
         $pk_P$ & Public key of the operator \\
         $sk_P$ & Private key of the operator\\
         $pk_D$ &\PaperName{} platform group public key\\
         $sk_D$ &\PaperName{} platform private group key\\
         $d_X$ &  Ephemeral private key of party X \\
         $K_s$ & Ephemeral session key \\
         $ATTR$ & Attestation Request \\
         $Q_V$ & Attestation quote from vSIM \\
         $|$ & Concatenation \\
         $[D]_k$ & Encryption of message D using key k \\
         $\sigma(D)_k$ & Signature of D with key k \\         
\hline

\end{tabular}
\end{table}

     \begin{figure}
    \centering
    \includegraphics[width=\columnwidth]{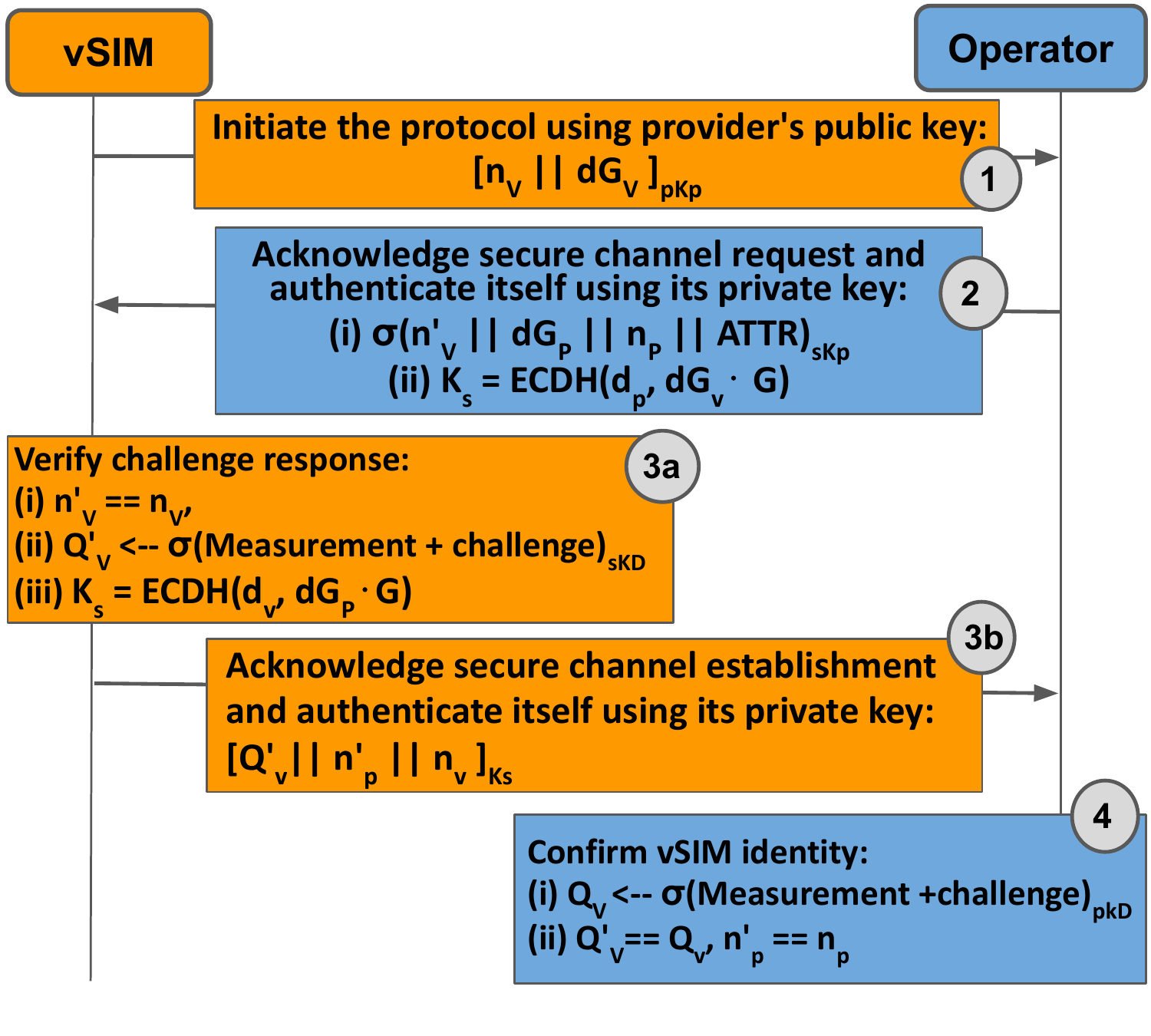}
    \vspace{-2em}
    \caption{Secure channel establishment protocol.}
    \label{fig:sec-channel}
    \vspace{-2em}
\end{figure}
\textbf{Secure Channel Establishment. }Before the operator is able to provision the device profile, it needs to ensure that it is talking to vSIM through a trusted channel. \PaperName{} assumes that the operator knows the valid state, i.e. the valid binary measurement of vSIM, and is able to verify its quote $Q_V$. 
The protocol notations are discussed in Table~\ref{tab:notation} and the process is shown in Figure~\ref{fig:sec-channel}. \bcircled[lightgray]{black}{1} The vSIM generates a random nonce $n_V$ as a challenge and a temporary Elliptic Curve (EC) private key $d_V$ and the corresponding public key $dG_V$, to derive a session key. $d_V$ remains securely stored within vSIM's protected memory throughout the session. This information is inaccessible by any untrusted software to maintain session integrity. vSIM then encrypts $n_V$ and $dG_V$ using the operator's public key $pk_P$ and transmits it to the operator. 

In the next step, \bcircled[lightgray]{black}{2} the operator receives and decrypts vSIM's request. The operator extracts $n_V$ and $dG_V$ using its own private key. The operator then generates its own random nonce, $n_P$, an ephemeral EC private key $d_P$, and its corresponding public key $dG_P$ and an attestation request vector $ATTR$. To authenticate itself to vSIM, the operator creates a signature of $n_V$, $dG_P$, $n_P$, $ATTR$, signed with $sk_P$ and sends it to vSIM. It is worth noting that there is no need to protect these four, because without the knowledge of $d_V$ or $d_P$, an attacker cannot gain or spoof any information. In addition, the operator derives the shared session key $K_s$ through the Elliptic Curve Diffie Hellman (ECDH) process~\cite{haakegaard2015elliptic}. 

\bcircled[lightgray]{black}{3a} When the information is received on the vSIM side, it verifies the operator's identity by comparing the sent and received nonces $n_V$. vSIM then invokes the attestor component (in the Security Monitor) with the challenge in the attestation request vector,  $ATTR$, to include it in the attestation response and to ensure its freshness. vSIM then uses $ATTR$ to generate the quote $Q_v$ signed by the \PaperName{} platform private group key $sk_D$, and then concatenates the quote with the nonces $Q_V || n_V || n_P$. Next, vSIM derives the shared session key $K_s$, through ECDH using its private key $d_V$ and the operator's public key $dG_P$. \bcircled[lightgray]{black}{3b} vSIM then encrypts this concatenated value using the derived session key $K_s$ and transmits it back to the operator. 
 
 \bcircled[lightgray]{black}{4} Finally, the operator decrypts the response using $K_s$ and extracts the vSIM quote $Q_v$ along with the nonces. The operator's Attestation Authority then validates the quote by recomputing the expected quote from the original challenge and \PaperName{}'s group public key $pk_D$, and comparing it with the received quote. If the quotes match, the operator verifies the freshness of the message by checking the included nonce values. Upon successful verification of both the attestation and freshness guarantees, the operator sends an acknowledgment to vSIM, announcing that the secure channel is established.

 \camera{The protocol is designed to provide mutual authentication, mutual non-repudiation, forward secrecy, freshness, confidentiality, integrity, and attestation-based trust assurance. Notably, although the protocol relies on the Attestation Authority to verify trusted execution and vSIM group membership, no device-specific identifiers are revealed during this process due to DAA. The provisioning party observes only the attestation quote, which reveals group membership but no individual device identity, and ephemeral key material that is discarded after session key derivation. Thus, even if compelled, the Attestation Authority cannot deanonymize a device, though it may deny service. If the provisioning party is compromised, forward secrecy ensures past sessions cannot be decrypted; however, the operator database mapping EIDs to IMSIs remains a target. But even in this case, since each subscriber profile is bound to a distinct device profile, the one-to-many mapping that enables cross-profile linkability is eliminated, and no correlation across subscriber profiles can be obtained.}








\subsubsection{Switching Between Profiles}

After establishing a secure channel, the operator sends a genuine device profile 
to the ME, which is handled by vSIM decrypting and securely storing it. To ensure the confidentiality and integrity of sensitive information such as authentication keys, attestation keys, subscriber identification, and other 5G security parameters, they are all stored using a secure storage device. The secure storage guarantees exclusive access for each trusted application to its own secure objects. This form of storage provides a private location for trusted applications, such as vSIM, to place their critical information. 

Once multiple device profiles are provisioned and securely stored within the device, \PaperName{} can switch between profiles without exposing an identity linkage. Profile Manager Proxy can do so by invoking \textit{switchDeviceProfile(EID)}, which will in turn disable the current profile and enable the one associated with the EID. Later, subscriber provisioning will happen using this device profile. As a simple proof-of-concept, we implement vSIM with the capability to store multiple profiles simultaneously and seamlessly switch between them. 
In the initial design, the vSIM software is responsible for isolating different profiles on the same device from one another. However, for further isolation, nested enclaves are the ideal architecture to isolate each profile in a child enclave of vSIM. This approach depends on the architecture of the TEE and whether it supports nested enclaves~\cite{9138984}. For TEEs that do support nested enclaves, each profile benefits from an additional layer of hardware-enforced memory isolation.

\subsection{5G Network Authentication}
\label{sec:arch:runtime}


vSIM aims to have the same security guarantees and operation as a physical SIM card once provisioned---since, at that point, the security model remains the same. The SIM functions as a root-of-trust for mobile communications needed for authentication by securely storing both the device profile and the subscriber profile information, such as the long-term secret key \textit{(k)} and the IMSI. This security property extends to the vSIM implementation, and here we overview how.

\textbf{Subscriber Profile Provisioning.} 
After provisioning a device profile (discussed in \S\ref{sec:arch:secchan}), the device can request a subscriber profile (the IMSI and long-term secret key) same as before, and the transmission of that information is what is considered to have a profile successfully provisioned. At this point, vSIM is equivalent to having a physical eSIM card.  
 
\textbf{5G Network Authentication.}
With a subscriber profile provisioned, the user can then connect to the network.  
When they do so, the 5G authentication process begins as specified in the 3GPP Specification~\cite{3gpp_ts_33_501}. Similar to the traditional process, the ME forwards the challenge to vSIM. Upon reception of the challenge, vSIM retrieves the long-term secret key (previously received from the operator) from its secure storage. Then vSIM generates the necessary cryptographic keys and a response. Once the keys and response are generated, vSIM transmits them back to the ME, which 
passes the messages to the network operator for verification---this interaction does not expose any more visibility of the response than exists today, and does not imply the ME needs to be trusted.   

\textbf{5G Network Connection and Traffic.} 
After successful authentication, the core network may occasionally request the IMEI, to which the UE can respond with a reprogrammed value (\S\ref{motivation:pattern}). Then, the phone is connected to the network and a vSIM (or any SIM) is no longer needed.  At this point the operator will have stored information about the state of that authenticated subscriber. \tamara{In addition, the operator will also have saved  the information about the subscriber that will allow it to identify traffic from that subscriber.  }
The software on the ME
stores the information in non-secure storage, and includes it in data traffic headers.

\section{Prototype}
\label{sec:prototype}

We leverage Keystone~\cite{lee2019keystone}, an open-source TEE based on the Rocketchip RISC-V architecture~\cite{Asanović:EECS-2016-17}, which serves as our chosen TEE foundation. The prototype is a co-design of hardware and software. We implement vSIM to be run as an application inside a secure enclave, and we extend the 5G infrastructure software for the rest of the \PaperName{} prototype.

\textbf{Hardware.}
We leverage the Rocket Chip, an open-source System-on-Chip generator, to extend the Berkeley Rocket Core, a synthesizable open-source RV64GC RISC-V core~\cite{Asanović:EECS-2016-17} to prototype \PaperName{}.  The chip generator is built using Chisel, a high-level hardware construction language derived from the Scala language that emits synthesizable RTL.  We leverage the Rocket Chip generator to generate the modified cores, caches, memory controller and interconnects to design an integrated System on Chip (SoC).  

\begin{figure} 
    \centering
    \includegraphics[width=\linewidth]{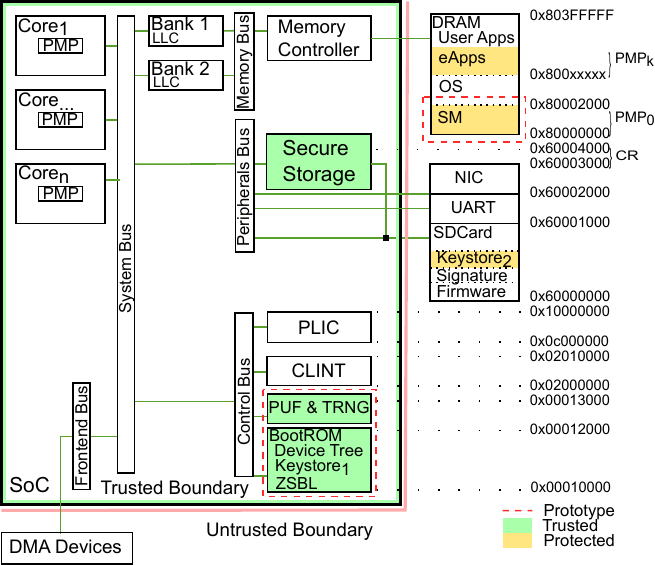}
    \caption{Hardware prototype and Memory-Map}
    \label{fig:vsim_hardware}  
    \vspace{-1em}
\end{figure}

We modify the original SoC design in three ways. First, we make changes to the memory controller to enable future work where it could perform the necessary features to incorporate secure memory,  encryption and hashing of data before it is sent/fetched to/from memory. Second, we add a secure boot-loader to the bootROM of the RISC-V design and update the security monitor
to leverage secure boot features, ensuring that cryptographic keys are only loaded in an authenticated and trusted environment. Third, we include the hardware-fused private key (in the form of a Physically Unclonable Function (PUF)) and a true random number generator (TRNG). These are used by the security monitor to derive an ephemeral key for the chain of trust during local attestation, making \PaperName{} unlinkable and unclonable.
Figure~\ref{fig:vsim_hardware} highlights the components that were either changed or added to the generic Rocket Chip SoC with dotted red rectangles. For the prototype, we store a cryptographic key (keystore 1) in the bootRAM logic; in a commercial system, this key would be stored in a secure storage device such as an eFuse memory region. The RISC-V Physical Memory Protection (PMP) tables~\cite{riscv_ismii} isolate memory-mapped regions at runtime. For example, the Security Monitor memory region is protected by PMP entry 0 and accessible only in machine mode.

\textbf{vSIM and Cryptographic Implementations.}
vSIM is implemented using the C language to run inside a Keystone enclave. Keystone relies on the Eyrie~\cite{lee2019keystone} security monitor to manage enclave applications. We port the Libsodium~\cite{libsodium} library into the enclave for software-based cryptographic operations. The prototype system uses \textit{X25519} elliptic curve cryptography, which is based on \textit{Curve25519} in Montgomery form, for key exchange~\cite{turan2019compact}. For digital signatures and quote signing, we use Ed25519, which generates 64-byte signatures~\cite{bisheh2021cryptographic}. 
The implementation uses the \textit{XChaCha20-Poly1305} IETF authenticated encryption algorithm, which combines the \textit{XChaCha20} stream cipher with the \textit{Poly1305} message authentication code~\cite{thara2020survey}. We use standard 24-byte nonces for each encryption operation. Each encrypted message includes a 16-byte authentication tag for integrity verification, with messages padded to fixed 32-byte boundaries to prevent length-based information leakage. All hash operations use SHA-256. 

Authentication, is implemented using the standard XOR operations as specified in 3GPP-TS 33.501 standard ~\cite{3gpp_ts_33_501}. MAC verification and SQN comparison are implemented using the Tiny-AES~\cite{tiny-aes} library in the enclave. In accordance with the standard \cite{3gpp_ts_33_501}, we implement the 5G authentication using Milenage algorithm suite with a 256-bit long-term secret key~\cite{yoo2019design}. 

Since it is not feasible to modify a commercial SM-DP+ server for profile provisioning, we emulate the server to implement the secure channel establishment. We implement it as a C++ based web server that contains a Remote Attestation Authority, which is responsible for verifying the attestation quote of vSIM. This Authority generates attestation challenges and keeps the valid state of vSIM to verify received quotes. For validation, it performs a comparison between the expected measurement hash and the received one. The server leverages the Libsodium~\cite{libsodium} library for cryptographic operations.

\textbf{5G Infrastructure.}  To verify that \PaperName{} seamlessly integrates with 5G infrastructure, we leverage srsRAN \cite{srsran_docs} along with srsUE as our prototype User Equipment (UE) and extend its source code with a new interface to communicate with the vSIM software instead of the USIM. In the absence of radio hardware support, srsUE can establish a connection with srsRAN using ZeroMQ \cite{zeromq}, a high-performance messaging library.

For the 5G Core Network (5GC), we employ Open5GS \cite{open5gs}, an open-source implementation of the 5G Core and Evolved Packet Core (EPC). srsRAN establishes a connection with the Access and Mobility Management Function (AMF) in the 5G Core network. The AMF is responsible for subscriber authentication, authorization, and mobility management.

\section{Evaluation}
\label{sec:eval}

With \PaperName{}, there are three key questions.  \textit{\textbf{Q1:}} does it provide suitable privacy enhancements, \textit{\textbf{Q2:}} is it compatible with 5G software and protocols, and \textit{\textbf{Q3:}} what, if any, overhead does this extra capability add.  After describing our experimental setup, we answer each of these questions.

\subsection{Experimental Setup} 
We use Xilinx Genesys2 FPGA board with 1GB 1800Mbps DDR3, to run the UE prototype.
We run srsUE in an untrusted Linux environment alongside vSIM, which is secured within Keystone TEE, the enclave. 
The srsRAN software and containerized 5G core network components are hosted on a personal laptop.
The board (in which the UE prototype runs) is connected to the srsRAN software and 5G core through an Ethernet connection (following srsRAN's support for transmitting radio samples over ZeroMQ instead of through an antenna, providing a consistent measurement environment~\cite{srs-zeromq}).
For profile provisioning operations, we run the server program in a C6220 Cloudlab~\cite{cloudlab} Node with an x86-64 architecture, 8 Intel E5-2650 CPU cores, and 65GB of RAM.

\subsection{Privacy}
\label{sec:eval:privacy}
 \begin{figure} 
    \centering
    \includegraphics[width=\linewidth]{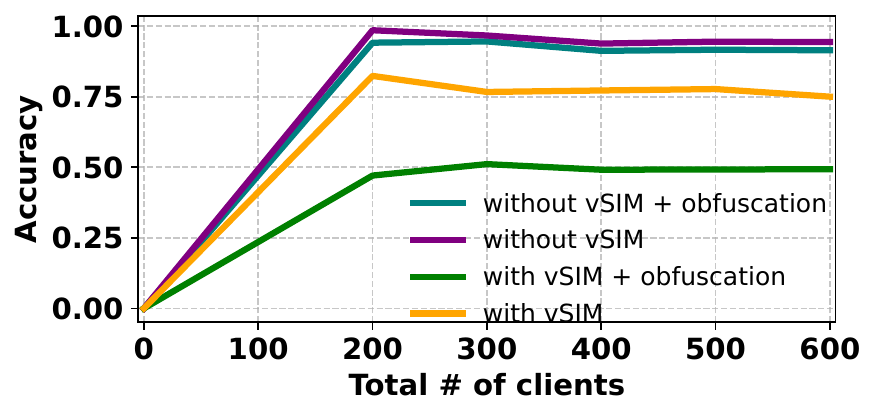}
    \vspace{-2.5em}
    \caption{SiamHAN model accuracy across different privacy configurations and client group sizes. The privacy configurations include the combinations of with and without vSIM, and with and without TLS pararameter obfuscation.}
    \label{fig:model}
   
\end{figure}

Recall that the attack \PaperName{} addresses is the operator's ability to correlate multiple subscriber profiles to perform pattern-of-life analysis---that is, tracking an individual's physical and online activities. In this section, we evaluate the effectiveness of \PaperName{} in thwarting this correlation.

5G networks are IP-based, meaning that the UE is assigned either a static or dynamic IP address, and all communication is carried over IP packets. An outside attacker can collect traffic to correlate connections; however, from the vantage point of a cellular operator (or multiple colluding operators), EID information is also available. \camera{In practice, an operator can deterministically correlate profiles through the EID. However, our goal in this experiment is not to model whether an operator (with full access) can perform correlation, but rather to isolate and quantify the contribution of the one-to-many EID mapping to that correlation capability.}

\camera{The ideal dataset would include a trace of a real 5G network, including both authentication and traffic, and would include users of the network switching eSIMs. This dataset does not publicly exist, to the best of our knowledge.  Instead, we use the strongest publicly available traffic-correlation model and augment its dataset with EID information. This also allows us to measure the additional linking power provided by EID when other deterministic identifiers are not available to the attacker (subscriber-level or IMEI-level defends discussed in \S \ref{motivation:pattern}), and to evaluate how effectively \PaperName{} removes that signal. As a baseline for comparison, we consider the case in which an individual uses different subscriber profiles but retains the same device profile (i.e., the traditional eSIM architecture) for different purposes or locations.}


In the SiamHAN paper~\cite{cui2021siamhan}, published at USENIX Security, \camera{the authors collect the CSTNET dataset and demonstrate the ability to correlate encrypted TLS traffic belonging to the same user across sessions with different IPv6 addresses using TLS handshake metadata and traffic characteristics.} Their algorithms do not inspect the EID (or any device profile, since they operate from the perspective of an outside attacker) yet still achieve 88\% accuracy (user discovery task with no prior knowledge). Using their publicly available code and dataset~\cite{siamhan-git}, we build upon their pre-trained model and modify the discovery task to measure how accurately it can diagnose users that re-appear on the network with a new node IP and correctly assign them to a previously seen user (i.e., correctly identify them). Accordingly, we define accuracy as the ability to correctly assign re-appeared users. We show that their model becomes significantly more effective in this task (93\% accuracy) when we incorporate EIDs into the ML model---reflecting a scenario in which the attacker is an operator with access to device profiles\footnote{\camera{A real operator knows the EID deterministically. The purpose of this experiment is to show the extra linking power EID gives when other identifiers are anonymized and how \PaperName{} eliminates that.}}


In our baseline experiment (without vSIM), we integrate a fixed EID into the SiamHAN model as an observable identifier available to the operator. In the \PaperName{} experiment, we integrate vSIM into the model, where each subscriber profile has its own unique device profile; when an individual changes subscriber profiles (which renews the IP address), the device profile changes as well---a property unique to vSIM.
This setup assumes that no traffic obfuscation is used. To investigate the impact of traffic obfuscation, we then add TLS obfuscation to both the baseline and vSIM experiments.




Figure~\ref{fig:model} illustrates the accuracy of the model for the identification task across different total numbers of clients. At each client population size, only a subset of clients return to the network, and the accuracy metric is computed solely over these returning clients. The results reveal a significant privacy improvement: the accuracy of correlating subscriber profiles in a network with \PaperName{} protected devices drops to 75\%, while the network without \PaperName{} achieves a much higher correlation accuracy of approximately 93\% (both without obfuscation).  When combined with obfuscation techniques, correlation accuracy drops dramatically to 49\% with vSIM---effectively randomly guessing---while without vSIM, the SiamHAN model remains highly resistant to the same obfuscation techniques, maintaining accuracy of 91\%. The detailed breakdown of aggregated correct versus incorrect re-identification of returning users is presented in Table~\ref{tab:metrics}, with 600 clients present in the network, 140 of whom return with a new IP.

\begin{table}[t]
\centering
\caption{Re-identification results for 140 returning clients (out of 600 total) that rejoin the network with a new IP.}
\vspace{-1em}
\label{tab:metrics}
\small
\setlength{\tabcolsep}{2pt}
\begin{tabular}{c|c|c|c}
\hline
 & \textbf{\textit{Correct}} & \textbf{\textit{Incorrect}} & \textbf{\textit{Accuracy}}\\
\hline \hline
\textit{w/o vSIM, w/ obfuscation}  & 128  & 12 & 91.4\% \\
\hline
\textit{w/o vSIM, w/o obfuscation}      & 130  & 10 & 92.9\% \\
\hline    
\textbf{\textit{w/ vSIM, w/ obfuscation}}      & \textbf{69}  & \textbf{71} & \textbf{49.3\%} \\
\hline
\textbf{\textit{w/ vSIM, w/o obfuscation}}      & \textbf{105}  & \textbf{35} & \textbf{75.0\%} \\
\hline\hline

\end{tabular}

\end{table}

\subsection{Compatibility}

We successfully validate that the UE can authenticate with the network and transmit data seamlessly through our FPGA prototype. When the UE initiates a connection to the network, it receives an authentication challenge from the 5G core to verify its identity. The UE then forwards this challenge to vSIM, which performs the necessary cryptographic calculations to generate the appropriate challenge response. This response is then returned to the UE, which forwards it to the network to complete the authentication process. After successful authentication, all subsequent connection and communication procedures continue according to standard protocols. This experiment shows that, the proposed architecture maintains compatibility with existing network infrastructure.

\begin{figure}
    \centering
    \includegraphics[width=1\linewidth]{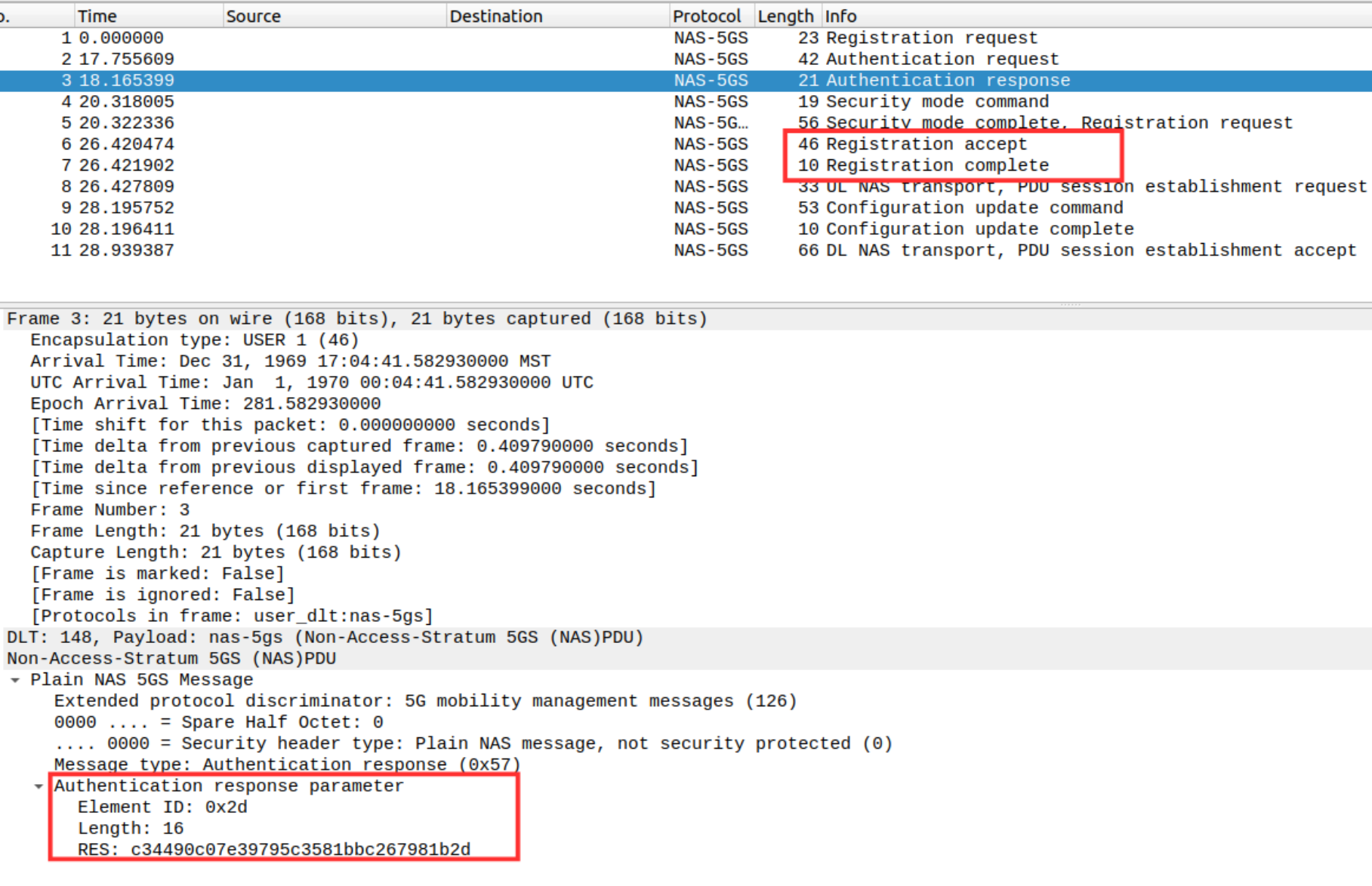}
    \vspace{-2em}
    \caption{Packets exchanged between vSIM and 5G core during UE successful authentication to the network. The red box 
    indicates the 
    response parameters calculated by vSIM.  }
    \label{fig:nas1}
  \vspace{-1.6em}
\end{figure}

Figure \ref{fig:nas1} illustrates the packet exchange between vSIM and the 5G core network during UE authentication. The bottom highlighted red box indicates the authentication response parameters calculated by vSIM and transmitted to the 5G core. Upon successful verification,
the 5G core completes the registration process by sending registration accept and registration complete messages to the UE.

\if{0}
\begin{figure}[ht]
    \centering
    \subfloat[]{%
        \includegraphics[width=\columnwidth]{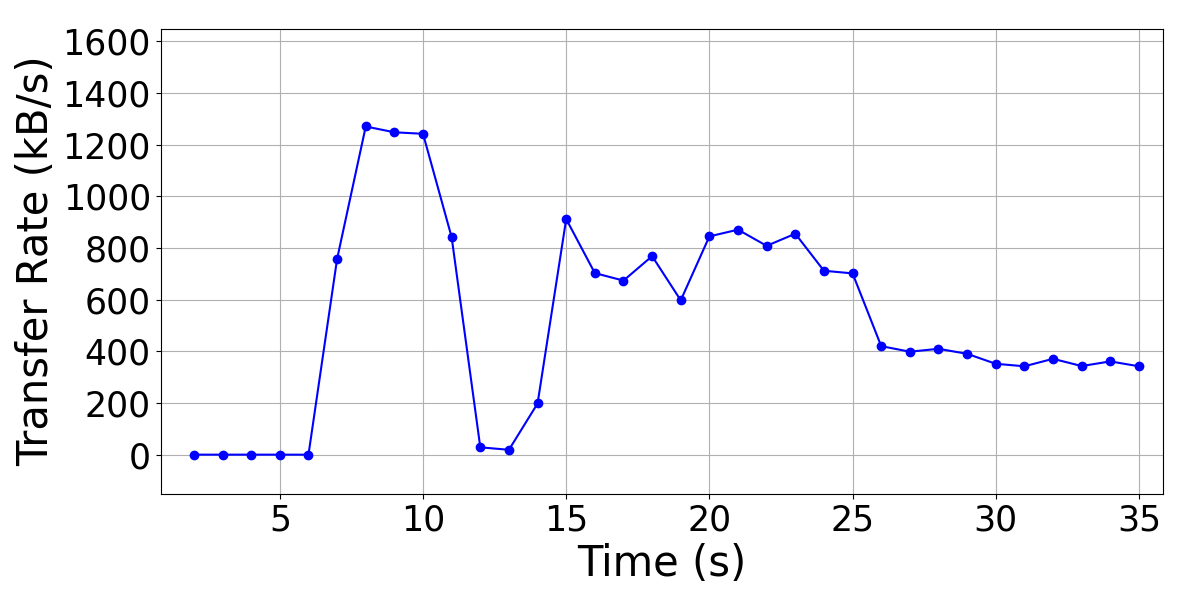}%

        }%
        \\
    \subfloat[]{%
        \includegraphics[width=\columnwidth]{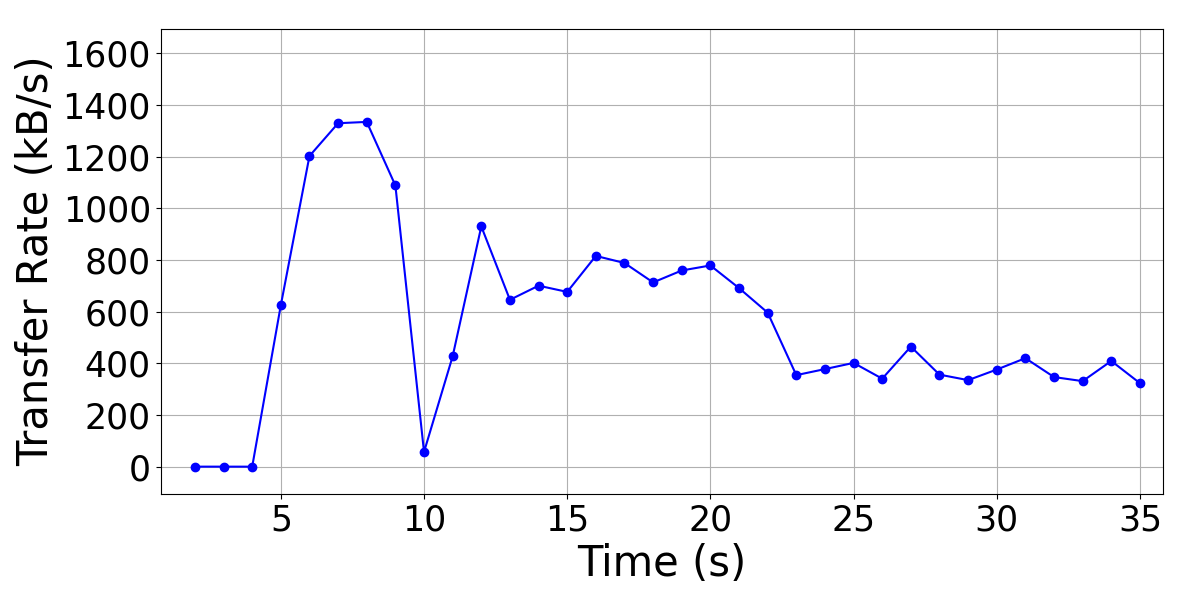}%
        }%
  
    \caption{Network usage of UE. (a) With vSIM, (b) Without vSIM}
            \label{fig:net}
\end{figure}
\fi

\begin{figure}
    \centering
    \includegraphics[width=1\linewidth]{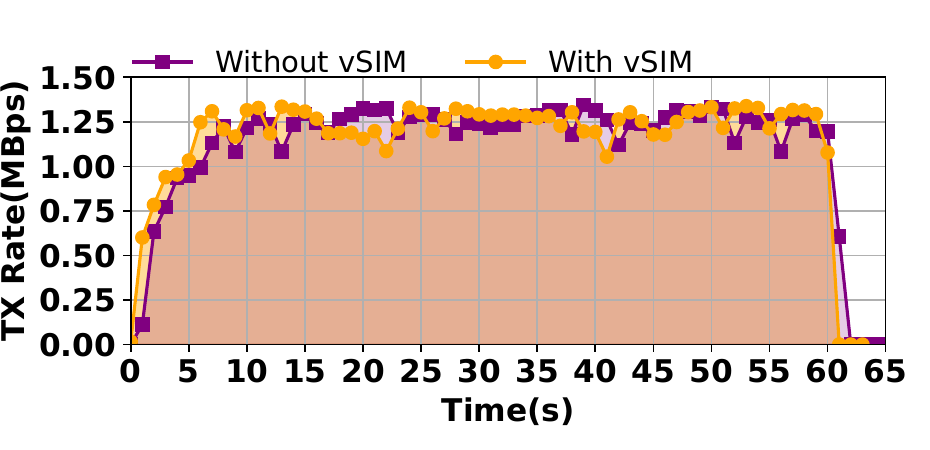}
    \vspace{-2.5em}
    \caption{Bandwidth usage of the device after authenticating the subscriber to the network with and without vSIM. }
    \label{fig:nas}
 \vspace{-1.8em}
\end{figure}

 The 5G authentication process is done only once when the UE wants to attach to the network, and therefore communication can occur without vSIM's involvement afterwards. To demonstrate, we setup an iperf3 server on the machine running the 5G core and a client on the \PaperName{} prototype (the UE). 
 The UE sends TCP traffic to the network both with using vSIM and without vSIM. The resulting measurements were identical (Figure~\ref{fig:nas}).
 


\textbf{Device Profile Provisioning.}
Device profile provisioning includes first establishing a secure channel.
We measured this taking 
approximately 805 milliseconds for the complete round-trip operation. This total time includes TCP/IP socket establishment and termination, message reception and parsing from the operator, message encryption and decryption operations, context switches between secure and normal execution, and digital signature generation and verification including quote processing. 
Of that 805 ms, the time required to prepare the quote on the device,
is approximately 19 ms. 
After the secure channel establishment, the device profile can securely be provisioned.  We measured 79ms for the operator to send a profile and vSIM to decrypt and store it in a data structure.
The device profile provisioning is considered a rare event, occurring at least in the order of days.

\textbf{UE Attachment to Network.}
Since vSIM runs inside an enclave, starting the enclave for each authentication adds latency. To address this overhead, we initialize the enclave when the UE starts. Our measurements show that computing the challenge response inside the enclave takes only 5 milliseconds. However, when integrated with the UE, the total time to transmit the challenge vector to vSIM and receive results through the shared memory channel increases to 658 milliseconds. For comparison, authentication without vSIM with an emulated eSIM takes about 7 milliseconds~\footnote{This latency is lower than what it would be with a real SIM card as the emulated eSIM only uses a function call to perform the read instead of actually reading the information from the SIM card.}. This increase can be attributed to the context switches between normal and secure execution, overhead of preparing data for secure communication, formatting the response vector between components, the execution of multiple communication transactions and computation functions to process the authentication vector. 
 \camera{However, authentication is an infrequent event in practice. 5G devices re-authenticate upon expiration of the registration update timer (typically on the order of hours) and during events such as coverage loss, airplane mode transitions, or device reboots~\cite{3gnas}. As a result, a smartphone is likely to perform only 10--30 authentications per day, depending on the user movement. Even in highly active IoT deployments, the rate is reported to be at most 48 authentications per day~\cite{iot}. At 658ms per authentication, this corresponds to approximately 6.6--20 seconds of daily overhead for a typical smartphone and about 32 seconds per day in the IoT case.}


In our implementation, switching between different subscriber profiles tied to different device profiles requires re-authentication of the profile to the network (multiple subscriber profiles can be provisioned simultaneously, but only one can be authenticated to the network at a time). 
Consequently, the latency of profile switching corresponds to the time required for the UE to deregister from the network, reinitialize the Non-Access Stratum (NAS) stack, and re-authenticate to the network. We measured this total profile switching time to be 1681.8 milliseconds. What this limits is the frequency of subscriber profile switching.  In our privacy evaluation from Section~\ref{sec:eval:privacy}, the profile switching occurs every 1 to 7 days.  As such, a 2s overhead is nominal. However, we see an opportunity to improve both the authentication time (through techniques to improve communication with the TEE, as in SafeBricks~\cite{safebricks}), and the profile switching time , through supporting multiple authenticated vSIM devices (through changes to the srsRAN UE codebase). 

However, switching between device profiles themselves, without switching between subscriber profiles, has a negligible latency of 11 ms. This time corresponds to disabling one profile and enabling another, which will later be used to provision a new subscriber profile.

\subsection{Performance}

\if{0}
\begin{table}
    \centering
    \caption{Latency comparison of authentication with and without vSIM in milliseconds}
    \begin{tabular}{|c|c|}
    \hline
         &  \textbf{Time (ms)} \\
    \hline
         \textbf{UE + vSIM} & 658.9   \\
         \textbf{UE + without vSIM}  & 7.6  \\
         \textbf{vSIM only}  & 5.2 \\
    \hline
    \end{tabular}
    \label{tab:auth}
    
\end{table}
\vspace{0.2in}
\fi


        
                


\textbf{Resource Usage Overhead.}
\begin{figure}
    \centering
    \subfloat[CPU usage.]{%
        \includegraphics[width=0.48\columnwidth]{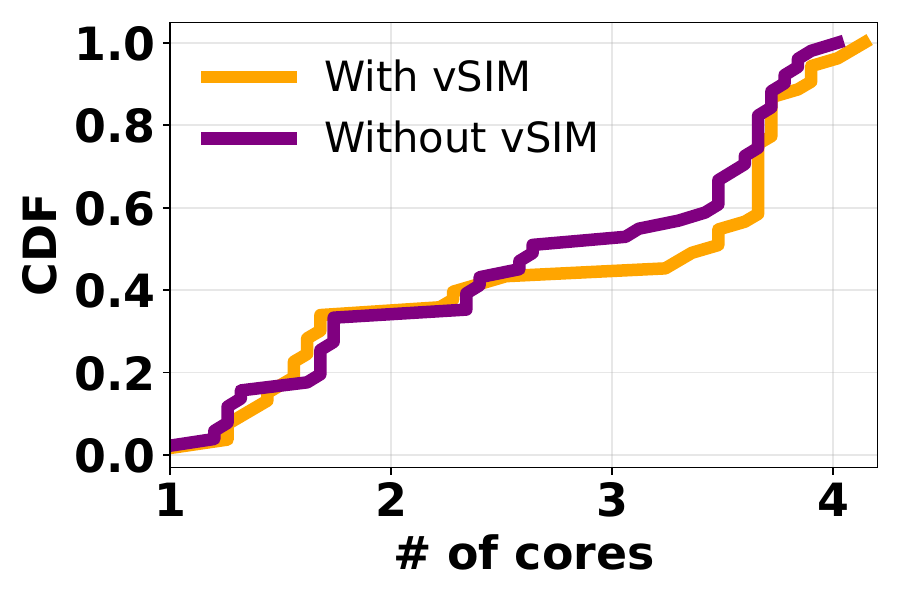}%
        \label{fig:cpu}%
    }%
    \subfloat[Memory usage.]{%
        \includegraphics[width=0.5\columnwidth]{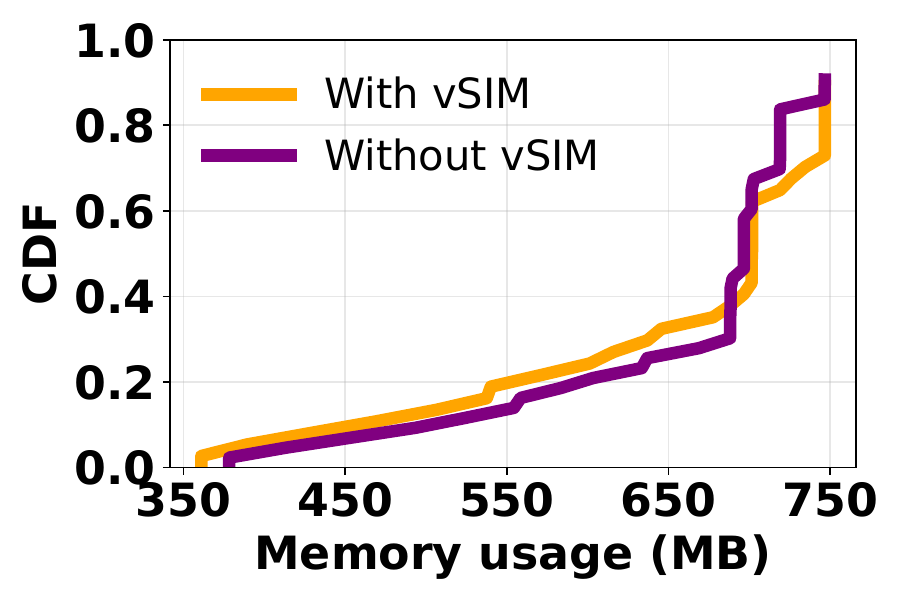}%
        \label{fig:mem}%
    }%
    \vspace{-1.0em}
    \caption{Resource usage CDF of UE with and without vSIM.}
    \label{fig:resource_usage}
    \vspace{-1.5em}
\end{figure}
As part of the resource analysis, we measure the CPU and memory usage for UE implementations with and without vSIM.  The experiments without vSIM are simply doing a function call emulating SIM functionality within the srsUE code.  We collect these performance metrics by 
leveraging the hardware counters on the board for both memory and CPU overtime. 
Figure \ref{fig:cpu} presents the prototype CPU usage as a cumulative distribution function (CDF).
These results show that there is no significant difference in CPU utilization between the two designs. Both experiments report that 50\% of the time less than 3 cores were used.



Figure~\ref{fig:mem} shows the prototype memory usage CDF
with and without vSIM.
The measurements reveal less than 1\% memory overhead when using vSIM compared to the baseline, which is attributed to the additional memory requirements of vSIM and the shared memory channels established for secure communication.

\if{0}
\begin{figure}
    \centering
    \includegraphics[width=1\linewidth]{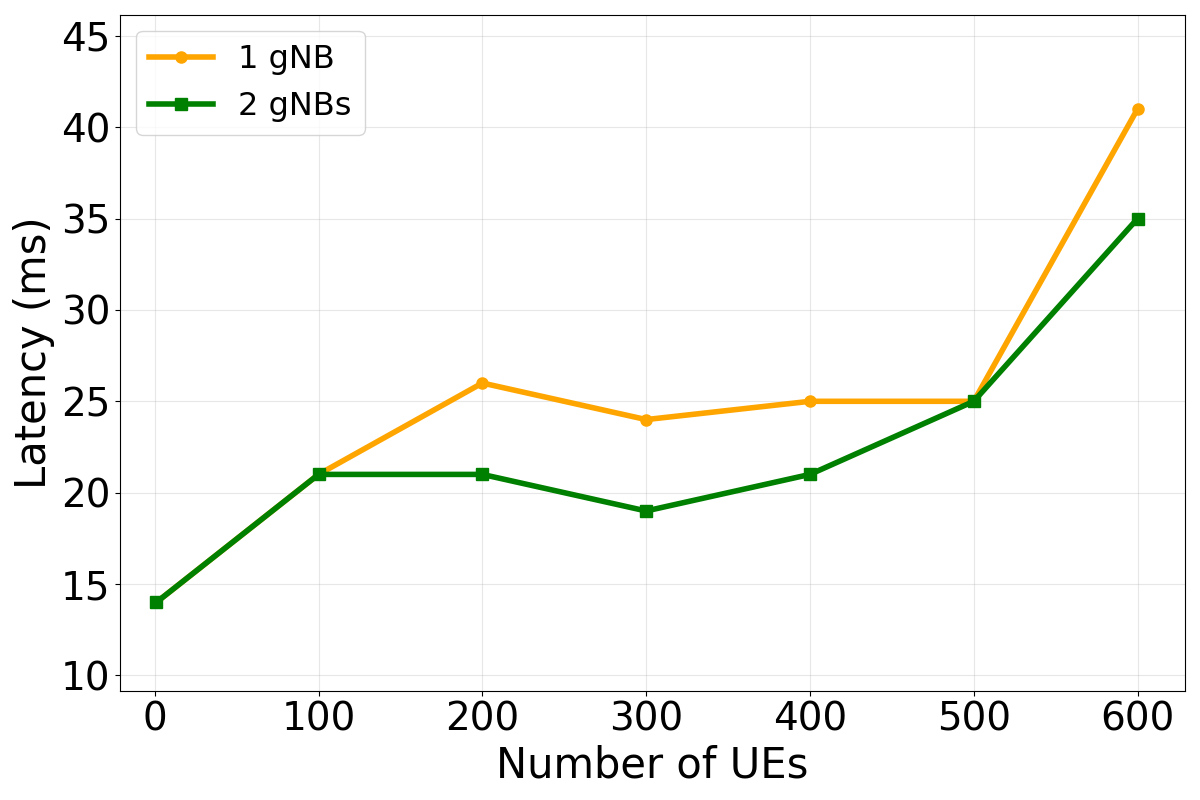}
    \caption{Latency of authenticating a UE by 5G Core with different number of concurrent requests. The orange line is using 1 gNodeB and the green line is using 2 gNodeBs with even loads.}
    \label{fig:infra-overhead}
\end{figure}
\fi

\section{Discussion}
\label{sec:discussion}

There are some open questions that are not addressed in this paper that impact the overall success of \PaperName{}.

\textbf{Adoption and Backwards Compatibility.}
The first is that \PaperName{} is introducing new capabilities into 5G networks that do not exist today.  As such, for it to have value, it requires adoption by mobile network operators.  We see this as a very viable proposition.  As discussed and demonstrated, authentication with a vSIM that has a device profile provisioned is fully compatible with existing 5G network infrastructure.  The difference is the dynamic provisioning of a profile. However, there is a path towards incremental adoption where device profiles can be provisioned through a Mobile Virtual Network Operator (MVNO) or a trusted third party, and after that, without any changes to the operators, vSIM can be used.


Then, for those operators that wish to take advantage of the dynamic device profile provisioning with vSIM, they can provide the device profile provisioning features themselves.  
%
%
From a business perspective, privacy is a desirable service that consumers are willing to pay for. This can be seen by the adoption of VPN technology, along with privacy preserving chat applications, like Signal~\cite{signal}.
Businesses with any travel into certain regions that might put their employees at risk, absolutely need privacy enhancing capabilities that protect the employees from being tracked.  
As such, \PaperName{} is a value-added service that mobile operators could charge for.


\camera{\textbf{Side-Channels and Correlation Attacks.}
By replacing a dedicated tamper-resistant chip with a TEE application on the main SoC, specific security trade-offs arise. \PaperName{} mitigates the most critical software-level concerns---including TCB integrity---through secure memory, remote attestation, and a verified boot chain. Nevertheless, TEEs remain more susceptible to side-channel attacks (e.g., cache-timing~\cite{cache}, power analysis~\cite{power}) than a discrete secure element, whose physical isolation and constrained execution environment offer a stronger boundary against such threats by design.

Beyond implementation-level side channels, there are also privacy risks that stem from the observable use of multiple device profiles over time. In this paper we do not address the situation where the adversary can track the location of each vSIM device profile.  
If an adversary sees two different device identifiers in the same location within some period, it provides evidence that they could
be the same individual (knowing that vSIM exists)--- specially in cases where ephemeral subscriber identity or rotation is used to provide location and identity privacy.   Examining other locations where the same device identifiers show up along with commonalities in traffic patterns, may give the adversary the ability to correlate the different identifiers.
With enough care in cycling through device profiles and using obfuscation technology, we believe it is possible to frustrate this effort.}



\textbf{Network Infrastructure Overhead.}
The core infrastructure difference between current 5G 
and \PaperName{} is that a device can switch 
profiles at runtime.  This feature results in extra authentications on the network.  The amount of extra load is determined based on the frequency of device profile switching.  In our privacy evaluation, where we demonstrated significant enhancements, we only switch once every 1-7 days.  This overhead is only a nominal additional load on the core infrastructure.  One of the future goals of \PaperName{} is to introduce the ability to have multiple, concurrently authenticated device profiles. This feature would reduce the need to continuously authenticate the new profile, and would result in even smaller overhead on the network.


 \section{Related Work}
\label{sec:related}


\camera{In this paper we introduce an architecture which decouples the device profile from the subscriber profile for 5G privacy. While prior work has implemented software-based SIMs or addressed subscriber-level privacy (discussed further below), \PaperName{} is the first to implement a TEE-backed dynamic device profile provisioning system that breaks EID linkability across subscriber profiles in 5G networks. There are several pieces of related work focused on different aspects of this problem, which we discuss in turn.}

\textbf{Software SIM and Hardware-Level Identity Decoupling.}
\camera{Most related to our system design is SIMurai~\cite{simuraei}, which also implements SIM functionality in software and integrates with srsRAN and srsUE for 5G connectivity. However, SIMurai is a security research platform designed to explore hostile SIM attack vectors against baseband firmware and it retains a fixed device identity and introduces no mechanism to provision or switch device profiles.}

A manual approach to solve hardware linkability problem is to physically swap out SIM cards (or phones) regularly or for specific purpose.  This approach is limited -- in number of virtual identities one might have, and practicality of when they can be swapped.  An extension of that concept is to use a device that has a pool of SIM cards for which a phone can forward authentication requests~\cite{oh-ndss2023, glocalme}.  These SIM Boxes, are bulky, limited in practicality, and require physical access.  With vSIM, 
we allow
multiple device profiles on the phone itself, limited only by storage on the device, and introduce an architecture which is able to preserve the trust relationship needed in 5G authentication.



\textbf{Vulnerabilities Due to Identifiers.}
In 5G, the IMSI is a globally unique number that uniquely  identifies a subscriber.  As such, this is a valuable target for attackers to try to determine or associate with some other information~\cite{blackhat-imsi-2014}.  One example seeks to track a user's location through various control signals in the 5G protocol~\cite{shaik-ndss2016}.  Another example demonstrates an ability to expose the user's IMSI and IMEI, and map a victim's soft-identity (e.g., phone number) to its device~\cite{hussain-ndss2019}.
These pieces of work 
tie the IMSI to a device enabling correlation across subscriptions. Hong \emph{et.al.} studies a similar problem with temporary identifiers~\cite{hong-ndss2018-guti}.  These prior pieces of work are orthogonal and complementary to our work, as they are studying the subscriber profile, not the device profile.   We allow for multiple device and subscriber subscriptions per device and further decouple the device from the subscriber.  






\textbf{Subscriber-Level Identity Decoupling.}
Following from the exposed vulnerabilities associated with identifiers in 5G networks, there is work in decoupling the subscriber information from the subscriber identifier.
Prior work lays out how to protect subscriber identities, focusing on how to reallocate and refresh these identifiers\cite{nist}. 
PGPP introduces the ability to decouple the SUPI / IMSI from the user identity~\cite{schmitt-usec2021}.
AKAA introduces an anonymous authentication and key agreement scheme designed to protect against mobile tracking, solving the problem of the IMSI and privacy~\cite{yu-ndss2024-aaka}. 
Another work, while not explicitly decoupling subscriber profile from identity, examines the subscriber profile provisioning process itself~\cite{ahmed2024}. Similarly, \camera{SecureSIM~\cite{securesim} proposes certificate-based authentication and fine-grained access control to protect SIM files and credentials from external attackers, and further encrypts the APDU channel using Diffie-Hellman key exchange. While it shares with \PaperName{} a move beyond PIN-based trust and secure channel design, it leaves the device identifier fixed and therefore does not address the cross-profile linkability problem that \PaperName{} directly targets.}





\textbf{Decoupling Application/Network-Level Identifiers.}
The notion of identifiers being tightly coupled is more general than just the 5G protocols.  At the network level, the IP address is an identifier that can make all traffic from that same IP address correlated.  Solutions include Onion routing~\cite{dingledine2004tor} and VPNs~\cite{alshalan2016vpnsurvey}, along with some more speculative solutions such as assigning random IP addresses~\cite{raghavan-pets2009}, and letting hosts generate temporary IPv6 addresses~\cite{rfc3041}.  
At the application level, web browsing can be tracked through cookies and advertisements.  
There are a number of pieces of work that defend against this type of tracking~\cite{roesner-nsdi2012, walls-hotsec12}.  Most related is pseudonyms~\cite{han2013pseudonyms}, which provides user controlled privacy through the ability to direct the browser to use different identities for different sites.  In our work, this control is at the device profile and subscriber profile level, allowing user controlled privacy through being able to switch between them.

\section{Conclusions and Future Work}
\label{sec:concl}
We present \PaperName{}, which virtualizes the SIM card and introduces an ability to dynamically provision device profiles.  This capability thwarts the ability to perform pattern-of-life analysis attacks, which we demonstrate \PaperName{} can reduce correlation accuracy from 93\% to 49\% when combined with obfuscation techniques.  We overcome the core challenges of moving away from an existing model dependent on the physical trust boundary of a SIM card
through a set of techniques to establish 
trust.
Our prototype implementation confirms that \PaperName{} remains fully compatible with current 5G infrastructure for network authentication. 

As part of the future work, we will extend 
\PaperName{} in several directions.
First, we plan to evaluate vSIM with software-defined radio hardware to verify real-world performance. This approach reduces CPU overhead compared to the current setup.
%
In addition, we plan on adding a profile migration capability 
on different devices. This additional feature will further enhance privacy, masking the individuals actual location.  
%
Finally, in our current implementation, switching a profile requires re-authenticating to the 5G network.  We will investigate ways to enable
having multiple active authenticated subscriber and device profile pairs. 
This would allow users,
 in the extreme case, to be able to interleave packets from different profiles to decrease linkability.

\begin{acks}
\camera{This work was supported by in part by NSF through the Convergence accelerator program (Award 2326835), and by NSF through the CNS Core program (Award 2241818). The authors used generative AI-based tools solely to revise the text, improve flow and correct any typos and grammatical errors.}



\end{acks}
\bibliographystyle{plain}
\bibliography{refs}

\end{document}